\documentclass{article}
\usepackage[utf8]{inputenc}
\usepackage[a4paper, total={6in, 8.5in}]{geometry}
\usepackage{graphicx}
\usepackage{todonotes}
\usepackage{authblk}
\usepackage{url}
\usepackage[nottoc,notlot,notlof]{tocbibind}
\usepackage{csquotes}
\usepackage{physics}
\usepackage{amsthm}
\usepackage[font=small,labelfont=bf]{caption}
\usepackage{subcaption}
\usepackage[normalem]{ulem}
\usepackage{tikz}
\usepackage{mathtools}
\usepackage{qcircuit}
\usepackage{pgf}
\usepackage{amsfonts}
\usepackage{tabularx,ragged2e}

\newcolumntype{L}{>{\hsize=.6\hsize \arraybackslash}X}
\newcolumntype{s}{>{\hsize=.4\hsize}X}

\usepackage{paralist}
\usepackage[utf8]{inputenc}
\usepackage{pgfplots}
\DeclareUnicodeCharacter{2212}{−}
\usepgfplotslibrary{groupplots,dateplot}
\usetikzlibrary{patterns,shapes.arrows}
\pgfplotsset{compat=newest}
\usepackage{microtype}
\usepackage{algorithm}
\usepackage[]{algpseudocode}
\usepackage{etoolbox}
\usepackage{tikzsymbols}

\usetikzlibrary{
backgrounds,
arrows,
arrows.meta,
positioning,
decorations.pathmorphing,
decorations.markings,
snakes,
shapes,
fadings}

\usepackage{hyperref}
\hypersetup{%
  colorlinks=false,
}

\newcounter{thm}

\newtheorem{definition}[thm]{Definition}

\newtheorem{problem}[thm]{Problem}

\usetikzlibrary{decorations.pathreplacing}

\makeatletter
\newcommand*{\algrule}[1][\algorithmicindent]{%
  \makebox[#1][l]{%
    \hspace*{.2em}
    \vrule height .75\baselineskip depth .25\baselineskip
  }
}

\newcount\ALG@printindent@tempcnta
\def\ALG@printindent{%
    \ifnum \theALG@nested>0
    \ifx\ALG@text\ALG@x@notext
    \else
    \unskip
    \ALG@printindent@tempcnta=1
    \loop
    \algrule[\csname ALG@ind@\the\ALG@printindent@tempcnta\endcsname]%
    \advance \ALG@printindent@tempcnta 1
    \ifnum \ALG@printindent@tempcnta<\numexpr\theALG@nested+1\relax
    \repeat
    \fi
    \fi
}
\patchcmd{\ALG@doentity}{\noindent\hskip\ALG@tlm}{\ALG@printindent}{}{\errmessage{failed to patch}}
\patchcmd{\ALG@doentity}{\item[]\nointerlineskip}{}{}{} 
\makeatother

\algnewcommand\algorithmicforeach{\textbf{for each}}
\algdef{S}[FOR]{ForEach}[1]{\algorithmicforeach\ #1\ \algorithmicdo}

\tikzset{>=latex}

\def\BibTeX{{\rm B\kern-.05em{\sc i\kern-.025em b}\kern-.08em
    T\kern-.1667em\lower.7ex\hbox{E}\kern-.125emX}}

\algnewcommand{\LeftComment}[1]{\Statex \(\triangleright\) #1}

\makeatletter
\tikzset{>=latex}

\def\BibTeX{{\rm B\kern-.05em{\sc i\kern-.025em b}\kern-.08em
    T\kern-.1667em\lower.7ex\hbox{E}\kern-.125emX}}

\newcounter{linearProgram}
\newcounter{algo}
\newcounter{protocol}

\newenvironment{constraint}[1][htb]{%
  \let\c@algorithm\c@algo
  \renewcommand{\ALG@name}{Constraint Program}
  \begin{algorithm}[#1]%
  }{\end{algorithm}
}
\newenvironment{algo}[1][htb]{%
  \let\c@algorithm\c@algo
  \renewcommand{\ALG@name}{Algorithm}
  \begin{algorithm}[#1]%
  }{\end{algorithm}
}
\newenvironment{protocol}[1][htb]{%
  \let\c@algorithm\c@algo
  \renewcommand{\ALG@name}{Protocol}
  \begin{algorithm}[#1]%
  }{\end{algorithm}
}
\algdef{S}[FOR]{ForEach}[1]{\algorithmicforeach\ #1\ \algorithmicdo}
\makeatother
\algblock{Input}{EndInput}
\algnotext{EndInput}
\algblock{Output}{EndOutput}
\algnotext{EndOutput}
\algnotext{EndIf}
\algrenewcommand{\algorithmicthen}{:}

\newcommand{\RN}[1]{%
  \textup{\uppercase\expandafter{\romannumeral#1}}%
}
\algblock[Name]{Parallel}{End}
\algblockdefx[Name]{Parallel}{End}%
[1]{\textbf{in parallel for} #1 \textbf{do}}%
{\textbf{await all}}

\title{Distributed Quantum Computing and Network Control for Accelerated VQE}

\author[1,2]{Stephen DiAdamo}
\author[1]{Marco Ghibaudi}
\author[1]{James Cruise}
\affil[1]{Riverlane, St Andrews House, 59 St Andrews Street, Cambridge CB2 3BZ, UK}
\affil[2]{Technische Universit\"at M\"unchen, Arcisstraße 21, 80333 Munich, Germany}

\begin{document}


\clearpage

\maketitle

\begin{abstract}
    Interconnecting small quantum computers will be essential in the future for creating large scale, robust quantum computers. Methods for distributing monolithic quantum algorithms efficiently are thus needed. In this work we consider an approach for distributing the accelerated variational quantum eigensolver (AVQE) algorithm over arbitrary sized -- in terms of number of qubits  -- distributed quantum computers. We consider approaches for distributing qubit assignments of the Ansatz states required to estimate the expectation value of Hamiltonian operators in quantum chemistry in a parallelized computation and provide a systematic approach to generate distributed quantum circuits for distributed quantum computing. Moreover, we propose an architecture for a distributed quantum control system in the settings of centralized and decentralized network control.
\end{abstract}
\newpage
\tableofcontents
\clearpage

\section{Introduction}
To execute large scale quantum algorithms on a quantum computer will require a quantum computer to have a large number of qubits to both perform error correction and computation. One path to creating quantum computers with many qubits is to construct a network of smaller-scale quantum computers and perform distributed computing amongst them. This scheme is known as distributed quantum computing or quantum multi-computing. Envisioned in IBM's road-map for scaling quantum devices is a plan to create quantum interconnect to network dilution refrigerators each holding a million qubits to create a massively parallel quantum computer \cite{gambetta_2020}. In any system that is converted from monolithic to distributed, a layer of communication complexity is added in order to perform distributed operations across devices. In the context of distributed quantum computing, the control system is tasked with handling the needed communication and potentially synchronization between devices. The specifics of the stack strongly depend on the architecture of the distributed system. 

Connecting smaller quantum computers is one way to gain more power out near term quantum devices. Another possibility is to use variational quantum algorithms \cite{peruzzo2014variational}. Variational quantum algorithms are hybrid classical-quantum algorithms: they leverage classical optimization together with reduced-depth quantum circuits to generate an approximate solution to a problem. One such example is the variational quantum eigensolver (VQE). VQE can be used in quantum chemistry to estimate the ground state energy of molecular chemical Hamiltonians. An implementation of VQE on existing quantum computers have been presented by the quantum group at Google  \cite{arute2020hartree}. 

A modified version of VQE called Accelerated VQE or $\alpha$-VQE has been specifically designed to  make best usage of the near-term quantum hardware. The $\alpha$ in $\alpha$-VQE represents the trade-off parameter between run-time -- which could be long for some variational algorithms -- and circuit depth \cite{wang2019accelerated}. In other words, $\alpha$ allows to fine tune the run time to cope with the limited coherence time of near-term quantum machines. $\alpha$-VQE uses a more efficient method for estimating the expectation value of a Hamiltonian than standard VQE, which is a quantum algorithm called Accelerated Quantum Phase Estimation or $\alpha$-QPE. $\alpha$-VQE replaces the expectation value estimation stage of standard VQE with $\alpha$-QPE, thereby potentially enhancing VQE when longer qubit stability is achieved.

In this work, we take these concepts and combine them to construct a method for running $\alpha$-VQE on a distributed system of quantum computers. We begin with a technical overview of the higher level concepts that are used throughout the project. Next, we begin to decompose $\alpha$-VQE. To estimate expectation values in VQE, an Ansatz state has to initialized. In Section \ref{sec:multipartite} we consider various approaches for distributing the Ansatz states over an arbitrary distributed quantum computer and we propose a method for distributing the circuits needed to perform the Ansatz initialization. In Section \ref{sec:network_control}, we describe two different architectures for performing distributed quantum computing and propose network control systems based on Deltaflow.OS.

\subsection{Summary of Contributions}

In this work, we construct a framework for performing accelerated VQE on a distributed system of quantum computers. Our framework requires two inputs: the number of qubits in a distributed collection of QPUs and circuitry needed to run VQE. As an outcome we produce a mapping of the monolithic system to a distributed system such that the Hamiltonian expectation estimation can be performed in a parallelized and distributed computation. Moreover, we design a control system architecture that can be used to execute the distributed quantum gate instructions. This process is not strictly confined to $\alpha$-VQE and many of the ideas can be adapted for VQE in its standard form (or when $\alpha = 0$) and potentially other variational quantum algorithms. The strategy for distributing qubits and scheduling can also be adapted for other types of quantum algorithms, not necessarily variational. 

\subsection{Related Work}

We expand on the work of the accelerated variational quantum eigensolver in \cite{wang2019accelerated}, extending to distributed quantum architectures. The method of decomposing quantum algorithms that we use in our framework has been explored in \cite{yimsiriwattana2004generalized, eisert2000optimal}, where a full example of decomposing Shor's algorithm was proposed in \cite{yimsiriwattana2004distributed}. Control systems for quantum computing have been proposed in \cite{tannu2017taming, fu2016heterogeneous}, but these do not discuss the control between networked quantum computers. In \cite{meter2006architecture}, a quantum multi-computer architecture optimized to perform Shor's algorithm is proposed. Here we consider splitting Ansatz states and use a distributed and parallelized approach for executing $\alpha$-VQE, but there are overlapping ideas in these listed works in terms of the requirements for networking quantum computers.

\section{Technical Prerequisites}
In this paper, we make use of theory from distributed quantum computing, software based control systems, distributed operating systems and dataflow programming schemes. In this section, we give a brief overview of each of these topics as a primer.

\subsection{Distributed Quantum Computing}

Distributed quantum computing is the act of processing quantum information on two or more distinct quantum computers to solve a single problem and combining the results to produce one output \cite{cirac1999distributed, van2007communication}. According to the Dowling-Neven law \cite{salamin2014schrodinger, hartnett2019new}, the number of usable qubits in a single, monolithic quantum computer, is growing steadily in a Moore's like fashion. To generate larger quantum systems, an orthogonal approach relies on connecting multiple quantum computers and use classical communication and entanglement to perform distributed quantum computing. Classical communication and entanglement allow application of multi-qubit gates across physically separated quantum computers referred to as the LOCC-ENTANGLE model in \cite{denchev2008distributed}.

There are various ways to perform multi-qubit gates across quantum computers. In \cite{meter2008arithmetic}, teleportation is at the base of the overall process. In particular, two forms of teleportation - qubit teleportation and gate teleportation -- between quantum devices are analysed. It is shown that teleporting qubits performs better than teleporting gates. Teleportation requires one entangled pair and two bits of classical communication. If the qubits are to be teleported back to their original location, this operation would need to be performed twice. The approach we use in this paper, instead, uses the results from \cite{yimsiriwattana2004generalized}. As it will be described in more depth in Section \ref{sec:distributing}, Yimsiriwattana et al. do not use teleportation at all but instead rely on one entangled pair and two bits of classical communication to perform a distributed control gate. 

An analysis of how to perform quantum algorithms over a networked distributed quantum computer has been presented in  \cite{beals2013efficient}. A network model is proposed such that a distributed quantum system can simulate circuits for monolithic quantum computers with a communication overhead of $O(\log^2 N)$, where $N$ is the number of qubits in the full system. In  \cite{meter2008arithmetic}, it is discussed how a linear network topology will perform adequately for the foreseeable future, but I/O bandwidth will be a more challenging problem to overcome.

Another form of distributed quantum computing is cloud-based quantum computing, with companies such as Amazon, Microsoft, IBM, and others each releasing their own versions of a cloud quantum computing service \cite{cuomo2020towards}. In this type of distributed quantum computing algorithm input from a client is sent to a server, the server executes the algorithm instructions and then sends the results back to the client. In this case, protocols such as universal blind quantum computation \cite{broadbent2009universal} can be performed. In Section \ref{sec:network_control}, we consider among our models a cloud based model with cooperating vendors.

Overall, the development of distributing quantum computers will be a promising path to increasing the size of quantum computers. Many network technologies for high speed, low-latency communication have been developed in other contexts and as we will see, can potentially be applied to distributed quantum computing.

\subsection{Software Control Systems of Quantum Hardware}

To perform the gate operations on the qubits in a quantum computer requires a system that can translate gate instructions to physical interactions with the qubits. Currently, quantum algorithms are generally written in terms of circuits of quantum gates. The circuits are designed with the assumption of noiseless qubits. The software takes these circuits as input, optimizes them, and converts them to a data format such that the control system controlling the qubits can execute the instructions. 

A first step into defining how such a system functions is to draft a model of the full stack of the quantum computer. Such an architecture has been proposed in \cite{fu2016heterogeneous}, defining the software and hardware stack for a quantum computer. Here a protocol is defined to convert the classical and quantum instructions to binary strings such that they can be executed at the machine level. This stack incorporates error correction into the model and injects the additional instructions between gate operations when needed. To execute the instructions,  hardware is in place that quickly reads the binary strings and then runs the optical control on the qubits. Such hardware is proposed in  \cite{hornibrook2015cryogenic}. Here cryogenic field programmable gate arrays (FPGAs) are incorporated into the control system architecture to manipulate semiconductor-based qubits. In \cite{cruise2020practical}, it is explained how moving the parts of the classical control of quantum system to lower-latency hardware like FPGAs can greatly benefit near-term quantum computing. 

An important part of this system stack is to be able to control the amount of messages being passed to the hardware. The number of messages can grow very quickly when error correction is considered, and the bandwidth of the system can be used up completely with just instructions for error correction. In  \cite{tannu2017taming} QuEST (Quantum Error-Correction Substrate), an architecture that delegates the task of quantum error correction to the hardware, for overcoming this is proposed. 

Software and hardware will have to work closely together in an highly optimized way in order to reduce the amount of instructions while performing them with as low a latency as possible. As quantum hardware technology improves and as more research towards quantum software deepens, this area of quantum computing will become central to executing quantum algorithms on large scale quantum hardware.

\subsection{Distributed Operating Systems}\label{sec:dist_os}

In general, an operating system (OS) is a system that manages the resources on a computer, such as the random access memory or the CPU, such that multiple programs can run simultaneously without undesirably interfering with each other. An OS also provides an interface between the user and the hardware. We will refer to the class of OSs that run on a single computer a centralized OS. A distributed OS is an OS that runs on a cluster or group of computers which are physically separated and connected via a network \cite{tanenbaum1995distributed}. To the user of a distributed OS, it should appear as if their programs are running on a centralised OS. More specifically, a distributed OS should behave as an ordinary centralised OS with the caveat that the programs could be running at any physical location which is not known to the user. 

A distributed OS can be deployed in multiple ways \cite[Chapter 8]{tanenbaum2015modern}. One way is to deploy the OS such that there is a distinction between the types of nodes in the distributed system, \enquote{nodes} meaning the computers in the cluster. The distinction is generally that there is one computer which controls the rest of the system and the controlled nodes follow all the commands of this \enquote{controller} unit. An alternative configuration is that the network connecting the computers in the cluster are connected via an internet and a set of internet protocols are used to request resources from the nodes in the cluster and perform inter-process communication. One can think of this as a client-server relationship \cite{tanenbaum2007distributed}. In this case, we call the operating system a network OS. The main difference between a distributed OS and a network OS is in a network OS, the user is aware that multiple systems are being use, albeit programs appear to be running on a single system. 

One can deploy their systems as a distributed OS or a network OS or as a hybrid of the two. When deploying a distributed operating system one needs to find a good balance of some key properties to ensure that the operating system is robust, efficient, and can be scaled up. For example, one can potentially make the system very robust by adding abundant inter-node communication, but this could make the system inefficient or can overload the processors with messages to process. 

In this work, we are focused on a specific system with a specific use case. These are systems that have classical control but have hardware that establish quantum entanglement and classical communication. We explore how one can design a distributed OS where the distributed part we focus on is a cluster of quantum computers. We take into account the two models, the client-server model over a entanglement network and the single controller model. We use a specific control system, namely Deltaflow.OS which we explain in Section \ref{sec:network_control}, to explore these two models in depth.  

\subsection{Dataflow Programming}\label{sec:dataflow}

Dataflow programming is a method of programming that uses a network flow, or a directed graph, approach for developing algorithms \cite{johnston2004advances}. The nodes in the network hold the logic of the program -- or are constant valued -- and the flow in the network represents the inputs to the next nodes in the flow which is then processed and output to the next node in the network until the program is complete. Generally, in dataflow programming the nodes run in parallel and asynchronously. In real implementations, the nodes run idly, waking when they receive input to process. When the program starts, some nodes are selected to initialize without input, triggering the start of the program. Commonly used hardware programming languages like Verilog and VHDL use dataflow programming as a paradigm.

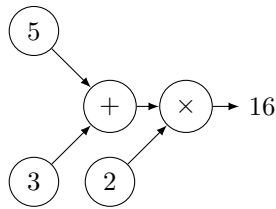
\begin{figure}[ht]
    \centering
    \begin{tikzpicture}
        \node[circle, draw=black, minimum size=.65cm] (n1) at (0, 0) {3};
        \node[circle, draw=black, minimum size=.65cm] (n2) at (0, 2) {5};
        \node[circle, draw=black, minimum size=.65cm] (n3) at (1, 1) {$+$};
        \node[circle, draw=black, minimum size=.65cm] (n4) at (1, 0) {2};
        \node[circle, draw=black, minimum size=.65cm] (n5) at (2, 1) {$\times$};
        \node[] (n6) at (3, 1) {16};
        \draw[->] (n1.north east) -- (n3.south west);
        \draw[->] (n2.south east) -- (n3.north west);
        \draw[->] (n3.east) -- (n5.west);
        \draw[->] (n4.north east) -- (n5.south west);
        \draw[->] (n5.east) -- (n6.west);
    \end{tikzpicture}
    \caption{An example of a dataflow program.}
    \label{fig:dataflow}
\end{figure}

In Figure \ref{fig:dataflow} is an example of a simple dataflow program. The constants $3$ and $5$ are inputs to the $+$ node which takes two inputs and outputs their sum. The output of the addition is passed to the $\times$ node, which takes two inputs and output the product. In this case, $2$ and $3+5$ are inputs to $\times$ and the complete output is $16$. 

In this work, we use Deltaflow to add control to the network hardware. Deltaflow is built on the dataflow programming paradigm. We will discuss in Section $\ref{sec:network_control}$ how Deltaflow can be used to define the logic in the control blocks for an overall network control.

\section{Distributing \texorpdfstring{$\alpha$}{a}-VQE}\label{sec:multipartite}
A problem to overcome when dealing with near-term quantum computing devices is that the ability to run deep circuits is greatly reduced due to low coherence time of qubit systems without error correction. A classical-quantum hybrid class of algorithms called \enquote{variational quantum algorithms} allow to run reduced depth circuits performing some of the algorithm on near-term quantum hardware and some on classical hardware. In particular, the variational quantum eigensolver (VQE) algorithm is a variational hybrid-quantum algorithm that can be used to find the minimum eigenvalue of a chemical Hamiltonian. It uses a quantum portion of the hardware to estimate the eigenvalues for a particular Ansatz of Pauli operations combining to form the Hamiltonian. VQE uses the quantum system to determine an expectation value and these expectation values are then combined to find an expectation value of the full Hamiltonian \cite{peruzzo2013variational}.

Using classical optimization techniques,  various Ans\"{a}tze -- plural of Ansatz -- are prepared with the goal of finding an estimate to the eigenstate with the lowest eigenvalue. The drawback of VQE is that the number of times the Ansatz state and expectation value needs to be prepared is proportional to $1/\epsilon^2$, where $\epsilon$ is the desired precision, which could lead to long run-times \cite{peruzzo2014variational}. Another way to estimate eigenvalues of unitary operations is using the quantum phase estimation (QPE) algorithm explained more in depth in Section \ref{sec:aqpe}. The advantage to using QPE is that the number of times the experiment is conducted to find the estimate is proportional to a constant. The downside is of course that the circuit depth grows proportionally to $1/\epsilon$.

As quantum hardware technologies improve, it will allow for longer coherence times of qubits and in turn allows for deeper quantum circuits. To make use of this ability, and to \enquote{squeeze} as much power out of the quantum hardware that is available, Wang et. al proposed the Accelerated VQE ($\alpha$-VQE) algorithm \cite{wang2019accelerated}. We again attempt to squeeze more power out of our quantum hardware by considering how one could implement $\alpha$-VQE for a distributed quantum computer. 

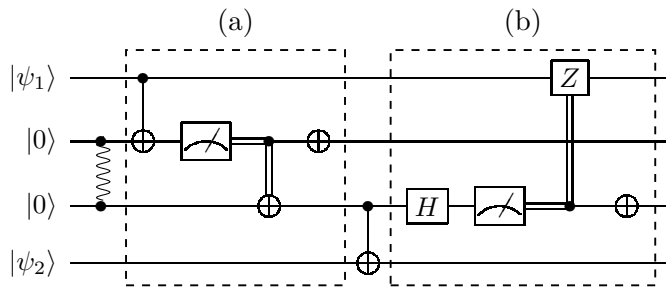
\begin{figure}[ht]
  \centering
  \begin{tikzpicture}[scale=1.2, every node/.style={transform shape}]
    \node (circ) at (0, 0) { 
      \scalebox{0.85}{
        \Qcircuit @C=1.0em @R=1.0em {
          \lstick{\ket{\psi_1}}    & \qw      & \ctrl{1} & \qw    & \qw       & \qw  & \qw     & \qw    & \qw  & \gate{Z} & \qw & \qw   \\
          \lstick{\ket{0}}         & \ctrl{0} & \targ    & \meter & \cctrl{1} & \targ  & \qw   &  \qw & \qw & \qw & \qw & \qw  \\
          \lstick{\ket{0}}         & \ctrl{0} & \qw      & \qw    & \targ     & \qw  &  \ctrl{1}   & \gate{H} & \meter & \cctrl{-2} & \targ & \qw \\
          \lstick{\ket{\psi_2}}    & \qw      & \qw      & \qw    & \qw       &\qw   & \targ     & \qw & \qw & \qw & \qw & \qw\\
        }
    }};
    \draw [dashed, line width=0.25mm] (-2.6, 1.3) rectangle (-0.2, -1.3);
    \draw [dashed, line width=0.25mm] (0.3, 1.3) rectangle (3.2, -1.3);
    \draw[snake=coil,segment aspect=0,segment length=4pt] (-2.85, -0.36) -- (-2.85, 0.35);
    \node[scale=0.9] at (-1.4, 1.6) {(a)};
    \node[scale=0.9] at (1.75, 1.6) {(b)};
  \end{tikzpicture}
  \caption{Circuit diagram for a non-local CNOT gate between $\ket{\psi_1}$ and $\ket{\psi_2}$ where (a) is the Cat-Entangler sequence and (b) the Cat-Disentangler sequence.}
  \label{fig:cat-entangler}
\end{figure}

When using VQE for quantum chemistry applications, it is common to prepare parameterized circuits that generate entangled Ansatz states. A commonly used Ansatz is the unitary coupled cluster Ansatz  \cite{romero2018strategies}, which grows in number of qubits required to prepare the Ansatz as Hamiltonian complexity increases. A critical part of using a distributed quantum computer for quantum chemistry is therefore preparing Ansatz states over an array of quantum computers. When distributing any quantum circuit across devices, the main complication that arises is when a controlled two qubit gate needs to be applied across two QPUs. There are two approaches we consider here. We assume that only entanglement and classical communication are used to achieve this. Alternative to this, we could consider physically moving qubits between QPUs but this is a much noisier task and we ignore this option. We consider two approaches: Teleporting one of the two qubits to the other QPU so that they are on the same QPU and then perform the two qubit gate on one QPU locally, the second approach is to use the mechanism introduced in Ref. \cite{yimsiriwattana2004generalized} where Yimsiriwattana et. al introduce \enquote{cat-entangle} and \enquote{cat-disentangle} protocols seen in Fig. \ref{fig:cat-entangler}. 

Comparing these two approaches in terms of number of operations needed, we find that using the approach of Yimsiriwattana et. al is more efficient. In order to use teleportation in a distributed system, we would require 2 Bell pairs to teleport the qubit from one QPU and back again. Using the method of Yimsiriwattana et. al requires just 1 Bell pair to perform a non-local control gate and this Bell pair can also be used to perform multiple control gates when the control qubit is the same as is done in \cite{neumann2020imperfect} for distributed quantum Fourier transform. 

Using the approach of Yimsiriwattana et. al, in the first subsection we consider how, given a collection of QPUs and an electronic molecular Hamiltonian, we can generate a schedule that can be used to estimate the expectation value of the Hamiltonian. We develop two approaches for solving this: the first is a greedy algorithm and the second uses constraint programming. In the next subsection, we consider how we can perform the needed $\alpha$-QPE step that is required for estimating expectation value over a distributed system and merge the ideas to produce a complete version of a distributed $\alpha$-VQE.

\subsection{Scheduling Hamiltonians}\label{sec:distributing}

An electronic molecular Hamiltonian $H$ can be written as a sum of a polynomial number (with respect to the system size) of Pauli matrices in the form of Eq. \eqref{eq:ham2}, where each $P_i \in \{I, \sigma_x, \sigma_y, \sigma_z\}^{\otimes n}$ is a tensor product of qubit $n$ Pauli operators (or the identify), called a Pauli string, and each $a_i \in \mathbb{R}$,
\begin{align}\label{eq:ham2}
  H = \sum_{i} a_i P_i. 
\end{align}

In order to more effectively use a networked quantum computer, we wish to use a parallelized and distributed approach to expectation value estimation. We motivate the approach as follows. Given the linear nature of estimating $\expval{H}{\psi}$, we can break up the summation into its pieces. We need to prepare an $n$ qubit Ansatz for each piece of the sum in order to estimate each $\expval{P_i}{\psi}$ independently to later rejoin the expectation values to estimate $\expval{H}{\psi}$. Given the distributed QPU architecture, we need to allocate the qubits in such a way that Ansatz states can be prepared for each $P_i$ in the sum. Later, the coefficients $a_i$ can be merged to produce a single value for $\expval{H}{\psi}$.

For Hamiltonians that require a large number of qubits, in this subsection, we consider methods that distribute the expectation calculation of the Pauli strings between a given distributed quantum computer. Here we model a collection of quantum processors \{QPU$_1$,...,QPU$_m$\} as a collection of $q_i \in \mathbb N$ qubits (respectively), all of which are located in the same device. Given a set of QPUs and a Hamiltonian in the form of a summation of Pauli strings, a distributed layout of the qubits with the required allocation of communication qubits is produced. 

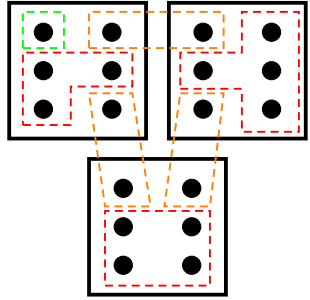
\begin{figure}
  \centering
  \begin{tikzpicture}[scale=0.6, every node/.style={transform shape}, 
      qpu/.pic = {
        \draw[line width=.5mm] (0, 0.85) rectangle ++(3, 3);
        \foreach \x in {0.75, 2.25} \foreach \y in {0.85, 1.70, 1.70 + .85}
        \draw[fill] (\x, 0.65 + \y) circle (2mm);
    }]
    \draw (0,0) pic {qpu};
    \draw (3.5,0) pic {qpu};
    \draw (1.75, - 3.45) pic {qpu};
            
    \draw [densely dashed, color=red, line width=0.25mm] (.3, 1.15) -- (1.35, 1.15) -- (1.35, 2.0) -- (2.7, 2.0) -- (2.7, 2.75) -- (0.3, 2.75) -- cycle;
    \draw [densely dashed, color=red, line width=0.25mm] (3.75, 1.95) -- (5.1, 1.95) -- (5.1, 1) -- (6.35, 1) -- (6.35, 3.65) -- (5.1, 3.65) -- (5.1, 2.75) -- (3.75, 2.75) -- cycle;
    \draw [densely dashed, color=red, line width=0.25mm] (2.1, -.75) rectangle (4.4, -2.4);
    
    \draw [densely dashed, color=green, line width=0.25mm] (.3, 3.65) rectangle (1.2, 2.85);
            
    \draw [densely dashed, color=orange, line width=0.25mm] (1.75, 3.65) -- (4.7, 3.65) -- (4.7, 2.85) -- (1.75, 2.85) -- cycle;
    \draw [densely dashed, color=orange, line width=0.25mm] (1.75, 1.85) -- (2.7, 1.85) -- (3.1, -.65) -- (2.1, -.65) -- cycle;
    \draw [densely dashed, color=orange, line width=0.25mm] (3.75, 1.85) -- (4.7, 1.85) -- (4.4, -.65) -- (3.4, -.65) -- cycle;
  \end{tikzpicture}
  \caption{Distribution of a 11-qubit Ansatz on three QPUs with 6 qubits each. One qubit is reserved for $\alpha$-QPE in green. Communication qubits are reserved in orange. The Ansatz qubits are in red. Two qubits are reserved for communication to accommodate for any control-control gates that could occur when running $\alpha$-QPE that need to cross QPUs.}
  \label{fig:distributed_hamiltonian}
\end{figure}

We enforce the following restrictions. Because the goal is to run $\alpha$-VQE, we know ahead of time that one additional qubit (additional to the qubits in the Ansatz) is reserved for each Ansatz to perform $\alpha$-QPE. Ontop of this, we need to reserve qubits for entanglement between QPUs which is necessary when an Ansatz is split between QPUs. The worst case for this occurs when there is a three qubit control gate (equivalent to a Toffoli gate) where the chain qubits are allocated on different QPU while performing $\alpha$-QPE.  In this case, since we are using the method of cat-entangling and disentangling, we need to reserve 2 qubits from each QPU for entanglement. We depict such a distribution in Figure \ref{fig:distributed_hamiltonian}. We formalize this as a problem:

\begin{problem}[Ansatz Distribution Problem]\label{prob:ansatz_dist}
Given a Hamiltonian $H = \sum_{i=1}^{n} a_i P_i$ where each Pauli string $P_i \in \{I, \sigma_x, \sigma_y, \sigma_z \}^{\otimes n_i}$ and a collection of $m$ QPUs described by the number of qubits on the system $[q_1, q_2, ..., q_m]$, output a series of rounds that can be used to estimate, for a given Ansatz $\ket\psi$, the expectation $\expval{H}{\psi}$. In order to prepare an Ansatz, when $P_i$ is split between two QPUs, 2 qubit from each QPU have to be allocated in order to perform non-local operations for preparing the Ansatz $\ket\psi$ across two or more QPUs. Moreover, 1 qubit needs to be reserved for $\alpha$-QPE. The solution to this problem outputs a schedule of distributions in which one can run over the distributed system to obtain an estimate to $\expval{H}{\psi}$.
\end{problem}

For the task of distributing the qubits, we take various approaches to this problem. In its essence, this problem is a resource allocation problem. We can therefore gain insight from common solutions to such problems. Common approaches for resource allocation problems are greedy algorithms and constraint programming. We propose an algorithm of each approach in this section. 

\begin{subsubsection}{Greedy Ansatz Distribution}\label{sec:ansatz_dist}
  In the greedy algorithm approach, we greedily fill the QPUs with as many Ansatz states that can possibly fit and for the remaining needed qubits, we split then across the QPUs reserving the needed qubits as needed. When the QPUs cannot fit any more Ans\"atze, the execution of those estimations are moved to the next round. In detail, we propose Algorithm \ref{algo:greedy}. We refer to an algorithm called \textbf{doesNotFit} which simply runs a similar logic as the main algorithm but just ensures a distribution exists for one particular Ansatz. We refer the reader to Appendix A, Algorithm \ref{algo:does_not_fit} for the detailed algorithm.
          
  \begin{algo}[H]
      \footnotesize
    \caption{Greedy Ansatz Distribution}
    \textbf{Input:} 
    \begin{compactitem}
      \item List of QPU sizes $Q = [q_1, q_2, ..., q_m]$.
      \item $n$ the qubits for Ansatz
      \item $p$ the number of Pauli strings to distribute
      \item Parameters for recursion defaulted to $schedule=\{\}$ and $round=1$
    \end{compactitem}
    \textbf{Output:} An Ansatz distribution schedule used to compute $\expval{H}{\psi}$ for an Ansatz $\ket\psi$ of size $n$ qubits. \\
    \textbf{GreedyDistribution}$(Q,n,schedule,round)$:
    \begin{algorithmic}[1]
      \If {$p=0$ or $n = 0$}
          \State return $schedule$
      \EndIf
      \State $Q' \gets \textbf{copy}(Q) = \{q_1', ..., q'_m \}$ \Comment{Copy $Q$ for modification}
      \State $schedule[r] \gets [\hspace{1mm}]$ \Comment{Initialize the schedule for this round}
      \State $couldNotFit \gets 0$
      \For {$i \in {1, ..., p}$}
          \State $\textbf{sort}(Q')$
          \If {\textbf{doesNotFit}($n, Q'$)}
              \If{$round = 1 \wedge i = 1$}
                  \textbf{exit} \Comment{The Ansatz does not fit, problem cannot be solved}
              \EndIf
              \State $couldNotFit \gets coundNotFit+1 $
              \State \textbf{continue}
          \EndIf 
          \State $distribution \gets [0 \text{ for } \underline{\hspace{2mm}} \in \{1,..,m\}]$ \Comment{A vector of $m$ zeros}
          \For{$j \in \{1, ..., |Q'|\}$}
          \State $curAllocation \gets [0 \text{ for } \underline{\hspace{2mm}} \in \{1,..,m\}]$
          \State $possibleQPUs \gets Q'|_{\{1, ..., j\}}$ \Comment{Restrict to the first $j$ available QPUs}
          \If{$j=1$} \Comment{No split  needed}
          \State $k\gets \textbf{QPUNumber}(possibleQPUs[1])$ \Comment{The QPU index}
          \State $curAllocation[k]\gets possibleQPUs[1]  - 1$
          \Else
          \State $k\gets \textbf{QPUNumber}(possibleQPUs[1])$ \Comment{The QPU index}
          \State $curAllocation[k] \gets possibleQPUs[1]  - 3$
          \For{$q'_s \in possibleQPUs|_{\{2, ..., j\}}$}
          \State $curAllocation[s] \gets q'_s - 2 $ \Comment{Reserve 2 qubits from the QPUs}
      \EndFor
      \EndIf
      \If{$\text{sum}(curAllocation) \geq n$} \Comment{An allocation is possible}
      \State $remaining \gets n$
      \State $iteration \gets 1$
      \For{$q'_s \in possibleQPUs$}
      \State $t \gets \min\{ remaining, curAllocation[s]\}$ 
      \State $distribution[s] \gets t$
      \State $remaining \gets remaining - t$
      \If {$iteration = 1$} \Comment{Remove the respective qubits from the first QPU}
      \If{$j = 1$}
      \State $q_s' \gets q_s' -  t - 1$
      \Else
      \State $q_s' \gets q_s' - t - 3$
      \EndIf
      \Else
      \State $q_s' \gets q_s' - t - 2$
      \EndIf
      \If{$remaining = 0$} \textbf{break}
      \EndIf
      \State $iteration \gets iteration + 1$
      \EndFor
      \State \textbf{break}
      \EndIf
      \EndFor
      \For{$q'_s \in Q'$}
      \If{$q'_s = 0$}
      \textbf{delete} $q'_s$
      \EndIf
      \EndFor
      \State $schedule[r].\textbf{add}((i, distribution))$
      \EndFor
      \State \textbf{return GreedyDistribution}$(Q, n, couldNotFit, schedule, round + 1)$ 
    \end{algorithmic}
    \label{algo:greedy}
  \end{algo}
\end{subsubsection}

\clearpage
\begin{subsubsection}{Constraint Programming Approach}
  As another approach to solving Problem \ref{prob:ansatz_dist}, we use constraint programming. The trade off with constraint programming is that setting up a collection of constraints is generally straight forward but solving constraint problems on a finite domain is generally NP-complete, trading simplicity for time. We construct the multi-objective constraint program in detail in Constraint Program \ref{algo:constraint_dist}. Using this constraint program repeatedly, we can produce a schedule by running the constraint program on the maximum number of Ans\"{a}tze that fit in the system and using a solution from the output, once per round, until all Ans\"atze are covered. 
  
  \begin{constraint}[H]
      \small
    \caption{Constraint Programming Distribution}
    \textbf{Input}:
    \begin{compactitem}
      \item $Q = [q_1, ..., q_n], \forall i, q_i \in\mathbb{N}$  a list of the number of qubits for each QPU in the system
      \item $A\in\mathbb{N}$ the number of qubits in the Ansatz
      \item $m\in\mathbb{N}$ the number of Ans\"{a}tze to fit
    \end{compactitem}
    \textbf{Variables}:
    \begin{compactitem}
      \item $x_{ij} \in \{0, ..., A\}$: The number of qubits from Ansatz $0 \leq i \leq m$ placed on QPU $j$
      \item $y_{ij} \in \{0, 1\}$: The QPE qubit for Ansatz $i$ on QPU $j$
      \item $z_{ijk} \in \{0, 2\}$: The number of qubits used to split Ansatz $i$ between QPUs $j$ and $k$
    \end{compactitem}
    \textbf{Objective Functions}:
    \begin{center}
      maximize $\sum_{ij} x_{ij}$, \text{minimize} $\sum_{ijk} z_{ijk}$
    \end{center}
    \textbf{Constraints}:
    \begin{enumerate}
      \item There's only one QPE qubit per Ansatz: 
            \begin{align*}
              \sum_{j=1}^n y_{ij} = 1, \hspace{2mm} \forall i \in \{1, ..., m\} 
            \end{align*}
      \item If the Ansatz is split, then both QPUs use qubits: 
            \begin{align*}
              z_{ijk} = z_{ikj}, \hspace{2mm} \forall i \in \{1, ..., m\}, j, k \in \{ 1,..., n \}, j \neq k, j < k 
            \end{align*}
      \item Ansatz is completely covered with one QPE qubit:
            \begin{align*}
              \sum_{j=1}^n x_{ij} + y_{ij} = A + 1, \hspace{2mm} \forall i \in \{1, ..., m\} 
            \end{align*}
      \item Qubits allocated do not exceed the number of qubits on the QPU. Note we can recycle the splitting qubits for multiple splits of the same Ansatz.
            \begin{align*}
              \sum_{i=1}^{m} x_{ij} + y_{ij} + \max_{\substack{k\in\{1,...,n\} \\ k \neq j}}z_{ijk} \leq q_j, \hspace{2mm} \forall j \in \{1, ..., n\}
            \end{align*}
      \item The Ansatz fits on one QPU or it is split:
            \begin{align*}
              \begin{aligned}
                & \max_{i \in \{1,..., A\}} x_{ij} = A \wedge \sum_{j=1}^{n} z_{ijk} = 0 \hspace{2mm}\vee                                             \\ 
                & \left|\{x_{ij} : j \in\{1,...,n\}, x_{ij} \neq 0  \}\right|  - 1 = |\{ z_{ijk} : j,k \in \{1,...,n\}, j\neq k, z_{ijk} = 2 \}| / 2, 
              \end{aligned}
              \forall i \in \{1, ..., m\}
            \end{align*}
      \item The QPE qubit exists on a QPU with Ansatz qubits:
            \begin{align*}
              \exists j\in\{1,...,n\} \hspace{1mm}  x_{ij} \neq 0 \wedge y_{ij} \neq 0,\hspace{2mm} \forall i \in \{1, ..., m\} 
            \end{align*}
    \end{enumerate}
    \label{algo:constraint_dist}
  \end{constraint}
\end{subsubsection}


\subsection{Distributing \texorpdfstring{$\alpha$}{a}-VQE} \label{sec:avqe}

As discussed in earlier sections, The variation quantum eigensolver (VQE) is a variational algorithm that uses a combination of quantum and classical components and can be used to estimate ground state energies in electric molecular Hamiltonians. To perform chemical calculations, VQE is used with a statistical sampling sub-routine to estimate expectation values with a given Ansatz with a classical optimizer to pick the parameters to minimize the expectation value. In Ref. \cite{wang2019accelerated}, a generalization of VQE is proposed, called $\alpha$-VQE. The generalization replaces the statistical sampling step with a subroutine called $\alpha$-QPE, which for the selection of $\alpha \in [0, 1]$ can behave as VQE does, but also can become more efficient by choosing $\alpha>0$, which requires the ability to run deeper circuits on quantum hardware.

In this section, we take the proposed $\alpha$-VQE in \cite{wang2019accelerated} and map it to a distributed system. The main theme in this section is applying non-local control gates over separated QPUs. We follow the approach of Refs. \cite{yimsiriwattana2004generalized, eisert2000optimal} using entanglement and classical communication to perform control gates across distributed systems, relying on the pre-allocated qubits from the previous section to hold the entanglement across devices.

\subsubsection{Distributing \texorpdfstring{$\alpha$}{a}-QPE}\label{sec:aqpe}

The quantum phase estimation (QPE) algorithm is an essential ingredient to many popular quantum algorithms -- one such being Shor's algorithm. First discussed by Kitaev in \cite{kitaev1997quantum}, QPE is used to estimate the phase of a quantum state $\ket{\psi}$ that appears after applying a specific unitary operation $U$ to it, where $\ket\psi$ is an eigenstate of $U$. Specifically, QPE aims to estimate the phase $\phi$ in $U\ket\psi = e^{2i\pi\phi}\ket\psi$ with high probability. In Fig. \ref{fig:local_qpe}, we depict a circuit representation of QPE applied to a qubit $\ket\psi$ where $n$ qubits are used to estimate $\phi$.

\begin{figure}[H]
  \centering
  \begin{tikzpicture}
  \node[scale=0.9] at (0, 0) {
    \Qcircuit @C=2em @R=1.25em {
      \lstick{\ket{0}} & \gate{H} & \gate{Z(M\theta)}  & \ctrl{1} & \gate{H} & \meter & \cw & {\hspace{4mm}E\in\{0,1\}} \\ 
      \lstick{\ket{\phi}} & \qw  {/_m}  & \qw  & \gate{U^M} & \qw &\qw   & \qw &
    }
  };
  \end{tikzpicture}
  \caption{Circuit diagram for RFPE. $Z(M\theta) \coloneqq \text{diag}(1,e^{-iM\theta})$.}
  \label{fig:rfpe_qpe}
\end{figure}
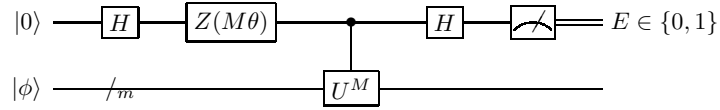

Here, we adapt a modified version of QPE developed in Ref. \cite{wang2019accelerated} called $\alpha$-QPE for a distributed system. $\alpha$-QPE is a modified version of rejection filtering phase estimation (RFPE) whose circuit diagram is given in Fig. \ref{fig:rfpe_qpe}. $\alpha$-QPE uses a free parameter $\alpha$ that is chosen depending on the available circuit depth on the specific hardware running the algorithm. With this $\alpha$, $M$ and $\theta$ are selected as $M = 1/\sigma^\alpha$ and $\theta = \mu -\sigma$. Here, $\sigma$ and $\mu$ are parameters for a normal $\mathcal{N}(\mu,\sigma^2)$ prior distribution in the first round of $\alpha$-QPE for sampling values of $\phi$, the \enquote{eigenphase} in $U\ket\phi = e^{\pm i \phi}\ket\phi$. Here $U$ is modified to be a rotation operator that rotates an Ansatz $\ket\psi$ by an angle $\phi$ in the plane spanned by $\{\ket\psi, P\ket\psi\}$, where $P$ is a Pauli string. More precisely, with the goal of estimating $|\expval{P}{\psi}|$, given an Ansatz preparation circuit $R\coloneqq R(\lambda)$ for some parameter vector $\lambda\in\mathbb{R}^n$ and a reflection operator $\Pi \coloneqq \mathbb{I} - 2\ketbra{0}$, $U \coloneqq R \Pi R^\dagger PR\Pi R^\dagger P^\dagger$ and the circuit depicted in Fig. \ref{fig:rfpe_qpe} is executed to obtain a value $E$. When $E$ is obtained, rejection sampling is performed to produce a posterior distribution, which can be shown to again be normal, in which to again sample values of $\psi$. This process is repeated until sufficient accuracy is reached. Once an estimate for $\phi$ is obtained, one can recover $|\expval{P}{\psi}|$ using the relation $|\expval{P}{\psi}|=\cos(\phi/2)$. In \cite{wang2019accelerated}, mechanisms to recover the sign of $\expval{P}{\psi}$ are provided.

In this subsection we tackle three key steps in to adapt $\alpha$-QPE for a distributed system: The first is mapping the state preparation circuit $R(\lambda)$ across multiple QPUs, the second is then to map $U$ to a distributed system, and the third, performing the controlled operation in Fig. \ref{fig:rfpe_qpe}. We solve these in order. The solution to the first task takes Ansatz preparation circuit $R(\lambda)$ and develops a mechanism such that it can be applied when some qubits are physically separated. Here we consider $R(\lambda)$ a variational form, a parameterized circuit used to prepare an Ansatz. We give an algorithm to achieve this in Algorithm \ref{algo:distributed_var_form}. 

\begin{figure}[H]
  \centering
  \subfloat[3 qubit variational form.]{
    \begin{pgfpicture}{0em}{0em}{0em}{0em}
      \color{red!50}
      \pgfsetlinewidth{.85pt}
      \pgfsetdash{{3.5pt}{2pt}}{0pt}
      \pgfrect[stroke]{\pgfpoint{-.85cm}{-0.4cm}}{\pgfpoint{9.75cm}{0.95cm}}
      \pgfrect[stroke]{\pgfpoint{-.85cm}{-2.55cm}}{\pgfpoint{9.75cm}{2.00cm}}
    \end{pgfpicture}
    \Qcircuit @C=1.125em @R=1.4em {
      \lstick{\ket{\psi_1}} & \gate{R_y(\theta_1)} & \gate{R_z(\theta_4)} & \ctrl{1} & \ctrl{2} & \gate{R_y(\theta_7)} & \gate{R_z(\theta_8)} & \qw  \gategroup{1}{4}{3}{5}{.9em}{--} \\
      \lstick{\ket{\psi_2}} & \gate{R_y(\theta_2)} & \gate{R_z(\theta_5)} &  \targ & \qw & \ctrl{1} & \gate{R_y(\theta_9)} & \qw  \\
      \lstick{\ket{\psi_3}} & \gate{R_y(\theta_3)} & \gate{R_z(\theta_6)} &  \qw & \targ & \targ & \gate{R_y(\theta_{10})} & \qw  \\
    }
  }
  \\[5mm]
  \subfloat[3 qubit variational form distributed across two devices.]{
    \begin{tikzpicture}
      \node (circ) at (0, 0) { 
        \scalebox{0.85}{
          \begin{pgfpicture}{0em}{0em}{0em}{0em}
            \color{red!50}
            \pgfsetlinewidth{.85pt}
            \pgfsetdash{{4.5pt}{3pt}}{0pt}
            \pgfrect[stroke]{\pgfpoint{-.85cm}{-1.2cm}}{\pgfpoint{15.0cm}{1.75cm}}
            \pgfrect[stroke]{\pgfpoint{-.85cm}{-4.0cm}}{\pgfpoint{15.0cm}{2.65cm}}
          \end{pgfpicture}
          \Qcircuit @C=1.0em @R=1.0em {
            \lstick{\ket{\psi_1}} & \gate{R_y(\theta_1)} & \gate{R_z(\theta_4)} & \ctrl{1} & \qw & \qw & \qw & \qw  & \qw & \qw  & \gate{Z} & \qw & \gate{R_y(\theta_7)} & \gate{R_z(\theta_8)} & \qw \gategroup{1}{4}{5}{12}{1.6em}{--} \\
            \lstick{\ket{0}} & \qw & \ctrl{0} & \targ & \meter & \cctrl{1}  & \targ  & \qw  &  \qw & \qw & \qw & \qw & \qw & \qw & \qw\\
            \lstick{\ket{0}} & \qw & \ctrl{0} & \qw & \qw & \targ &  \ctrl{1} & \ctrl{2}  & \gate{H} & \meter & \cctrl{-2} & \targ & \qw & \qw & \qw\\
            \lstick{\ket{\psi_2}} & \gate{R_y(\theta_2)} & \gate{R_z(\theta_5)} & \qw & \qw & \qw & \targ & \qw & \qw & \qw & \qw &  \qw & \ctrl{1} &\gate{R_y(\theta_9)} & \qw \\
            \lstick{\ket{\psi_3}} & \gate{R_y(\theta_3)} & \gate{R_z(\theta_6)} & \qw & \qw & \qw & \qw & \targ & \qw & \qw  &  \qw & \qw & \targ & \gate{R_y(\theta_{10})} & \qw\\
          }
      }};
      \draw[snake=coil,segment aspect=0,segment length=4pt] (-3.58,0) -- (-3.58,0.75);
    \end{tikzpicture}
  }
  \caption{An example of running the \textbf{DistributedRemapper} algorithm.}
  \label{fig:distributed_ansatz}
\end{figure}
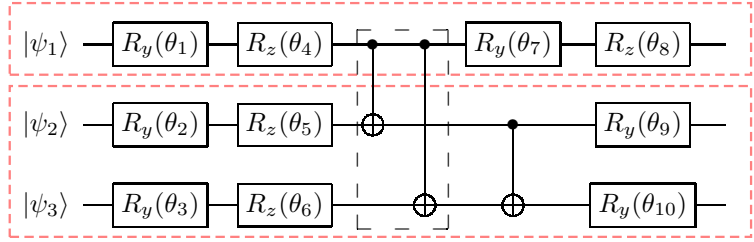
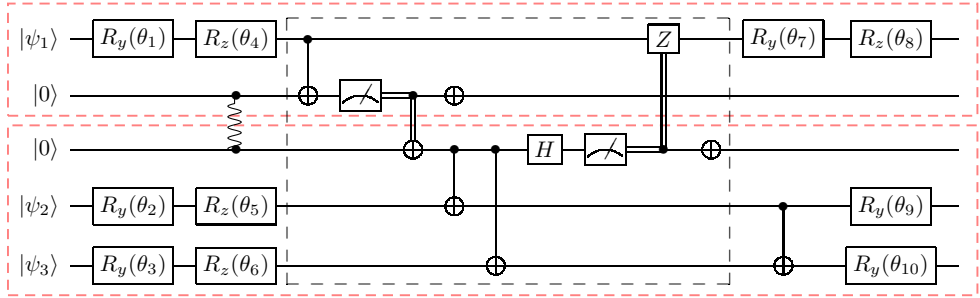

The high level idea of Algorithm \ref{algo:distributed_var_form} is, given the circuit representation of $R(\lambda)$ as a series of layers, where each layer is a collection of gates in a layer of circuit, and a mapping of qubits, to search for any control gates where the control and target are physically separated between two QPUs. When found, insert, between the current layer and next layer in the circuit, the necessary steps to perform the control gate in a non-local way using the cat-entangling method. We also ensure that entanglement is established between the two QPUs ahead of time by pre-pending an entanglement generation step. As an optimization, the cat-disentangler step can be shifted to a later layer if the non-local control gate has the same control qubit and no operations on that control qubit in between controlled gates.  Note that we can generate a distributed $R(\lambda)^\dagger$ in the same way. From the previous subsection, the proposed solutions to Problem \ref{prob:ansatz_dist} ensure that there are two qubits reserved on each QPU for the entanglement qubits needed for non-local operations. Producing the layering of a circuit can be done in a straight forward way and we assume that this structure is the input to the algorithm. We depict an example of running the algorithm in Fig. \ref{fig:distributed_ansatz}.

\begin{figure}[H]
  \centering
  \begin{tikzpicture}
  \node[scale=0.8] at (0, 0) {
    \Qcircuit @C=2.0em @R=1.0em {
      \lstick{\ket{0}} & \gate{H} & \qw & \qw & \rstick{\cdots} \qw & & \ctrl{4} &\multigate{3}{QFT_n^{-1}} &\meter & \cw   \\  
      & \vdots & & & & & & & \vdots \\
      \lstick{\ket{0}} & \gate{H} & \qw & \ctrl{2} &  \rstick{\cdots} \qw & & \qw & \ghost{QFT_n^{-1}} &\meter & \cw \\
      \lstick{\ket{0}} & \gate{H} & \ctrl{1} & \qw &  \rstick{\cdots} \qw  & & \qw & \ghost{QFT_n^{-1}} &\meter & \cw\\
      \lstick{\ket{\psi}} & {/_m} \qw & \gate{U^{2^0}} & \gate{U^{2^1}} &  \rstick{\cdots} \qw & & \gate{U^{2^{n-1}}} & \qw & \rstick{\ket{\psi}} \qw
    }
  };
  \end{tikzpicture}
  \caption{Circuit diagram for QPE with unitary operation $U$ and eigenstate $\ket\psi$.}
  \label{fig:local_qpe}
\end{figure}
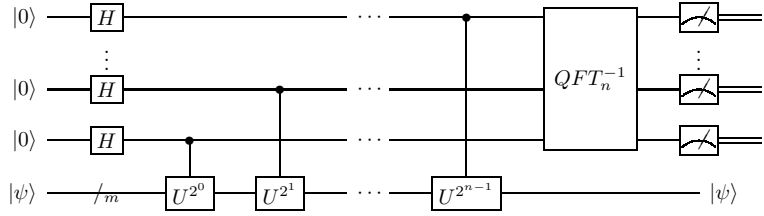

\begin{algo}[H]
    \small
  \caption{Local to Distributed Circuit}
  \textbf{Input:} 
  \begin{compactitem}
    \item A circuit representation of unitary $U$ where $U$ is a list of list of gates. Each list represents a layer in the circuit. Gates have the form $Gate(\text{ID})$ or $CONTROL(G,\text{ID}_1, \text{ID}_2)$ where $G$ is the gate to applied using control qubit with $ID_1$ and target qubit $ID_2$. The ID in the form $(i,j)$ where $i$ is the QPU and $j$ the qubit on that QPU.
    \item A qubit layout map $qubitMap$ on a collection ID tuples of the form $(ID_1, ID2)$.
  \end{compactitem}
  \textbf{Output:} An equivalent circuit that accommodates for non-local controlled gates. \\
  \textbf{DistributedRemapper}$(U, qubitMap)$:
  \begin{algorithmic}[1]
    \State $remappedCircuit \gets [[\hspace{1mm}]]$ \Comment{Add a placeholder layer in case an extra first layer is needed}
    \For{$layer_l \in U$}
    \State $modifiedLayers \gets [[\hspace{1mm}] \times 8]$
    \For{$gate \in layer_l$}
    \If {$gate$ \textbf{is} $CONTROL$} 
    \State $((i,j),(s,t)) \gets gate.\textbf{qubits}$ 
    \If{$i = s$}
    \State $modifiedLayers[0].\textbf{add}(gate)$ \Comment{Same QPU, no need to non-localize}
    \State \textbf{continue}
    \EndIf
    \If{ Entanglement is not established between QPUs $i$ and $s$}
    \State $remapping_{l-1}$.\textbf{add}($Ent((i, e_1), (s, e_2))$ \Comment{Add ent. gen. as previous layer}
    \EndIf
                    
    \State $modifiedLayers[0]$.\textbf{add}($CNOT((i,j), (i,e_1))$)
                    
    \State $modifiedLayers[1]$.\textbf{add}($c_i \gets measure((i,e_1))$)
    \State $modifiedLayers[2]$.\textbf{add}($c_s \gets classicalCommunication(i, s, c_i)$)
    \State $modifiedLayers[2]$.\textbf{add}($classicalCtrlX(c_i, (i, e_1)$)
    \State $modifiedLayers[3]$.\textbf{add}($classicalCtrlX(c_s, (s, e_2)$)
                    
    \State $modifiedLayers[4]$.\textbf{add}($Control-G((s,e_2), (s, t)$)
    \State $n\gets 0$
    \State  $\rhd$ Get series of control gates with same control qubit, target qubits on QPU $s$
    \For{$seriesGate \in$ \textbf{GetSeriesCGates}($U, layer_{l}, s, (i,j)$)}
    \State $n\gets n+1$
    \State $((\underline{\hspace{2mm}},\underline{\hspace{2mm}}),(\underline{\hspace{2mm}},t')) \gets seriesGate.\textbf{qubits}$ 
    \State $modifiedLayers.\textbf{add}([\hspace{1mm}],4+n)$ \Comment{Add empty list at index $4+n$}
    \State $modifiedLayers[4+n]$.\textbf{add}($Control-G'((s, e_2), (s, t')$)
    \State \textbf{remove} $seriesGate$ \Comment{No need to distribute in next iterations of parent loop}
    \EndFor
    \State $modifiedLayers[5+n]$.\textbf{add}($H(s,e_2)$)
    \State $modifiedLayers[6+n]$.\textbf{add}($c_s \gets measure(s, e_2)$)
    \State $modifiedLayers[6+n]$.\textbf{add}($classicalCtrlX(c_s, (s,e_2))$)
    \State $modifiedLayers[6+n]$.\textbf{add}($c_s \gets classicalCommunication(s, i, c_s)$)
    \State $modifiedLayers[7+n]$.\textbf{add}($classicalCtrlZ(c_i, (i, e_1)$)
    \Else
    \State $modifiedLayers[0].\textbf{add}(gate)$
    \EndIf
    \EndFor
    \State $remappedCircuit$.\textbf{addAll}$(modifiedLayers)$\Comment{Assume empty layers are ignored}
    \EndFor
    \State \textbf{return} $remappedCircuit$
  \end{algorithmic}
  \label{algo:distributed_var_form}
\end{algo}

\begin{algo}[H]
    \small
    \caption{GetSeriesControlGates}
  \textbf{Input:} 
  \begin{compactitem}
    \item $U$ the circuit as described in Algorithm \ref{algo:distributed_var_form}
    \item $layerl_l$ the current layer to decompose in $U$
    \item $s$ the QPU for the target qubit
    \item $(i,j)$ the control qubit
  \end{compactitem}
  \textbf{Output:} The series of gates directly following from layer $l$ that are control gates with with control qubit $(i,j)$. \\
  \textbf{GetSeriesCGates}$(U, layer_l, s, (i,j))$:
  \begin{algorithmic}[1]
    \State $layers \gets \{layer_{l+1},..., layer_{n}\} \subseteq U$ \Comment{Skip current layer} 
    \State $gates \gets [\hspace{1mm}]$
    \For{$layer_s \in layers$}
    \For{$gate \in layer_s$}
    \If{$gate$ \textbf{is} $CONTROL$ and $gate = (\underline{\hspace{2mm}}, (i,j),(s,\underline{\hspace{2mm}}))$}\Comment{Same control and target qubit}
    \State $gates.\textbf{add}(gate)$ 
    \Else
    \State \textbf{return} $gates$
    \EndIf
    \EndFor 
    \EndFor 
    \State \textbf{return} $gates$
  \end{algorithmic}
\end{algo}

The next step is to map $U \coloneqq R \Pi R^\dagger PR\Pi P^\dagger R^\dagger$ to a distributed system. One observation that can be made immediately is, since $P$ is a Pauli string, $P^\dagger = P$, so there are no additional steps needed to map $P^\dagger$. $P$ is a separable operation (i.e. there are no 2 qubit gates) and therefore we can apply each piece of $P$ in a single layer with no added inter-QPU communication. For mapping $R(\lambda)^\dagger$ to a distributed system, as discussed, given an $R(\lambda)$ as a circuit that is not distributed, we can obtain $R(\lambda)^{\dagger}$. To obtain the mapping, we can run Algo. $\ref{algo:distributed_var_form}$ with $R(\lambda)^{\dagger}$ as the input with the same Ansatz distribution. Next, we consider the $n$ qubit reflection operator $\Pi$ which can be decomposed (locally) as a series of single qubit gates and $CNOT$ operations. We can therefore again use Algo. \ref{algo:distributed_var_form} to map a provided reflection $\Pi$ to a distributed architecture given the Ansatz distribution as input.

For the control part of $\alpha$-QPE, we consider the controlled version of $U$, $c-U$, because of the structure of $U$, one can see that the only operation that is in need of control is in the reflection $\Pi$ since if $\Pi$ is not applied, $c-U$ is reduced to the identity. Here it will be the case that we need to execute control-control gates (CC-gates). If the Ansatz is split between QPUs, then two qubits need to be reserved on each QPU to accommodate for CC-gates. This is guaranteed by the scheduling algorithm in the previous subsection and there will always be two free qubits reserved such that we can apply Algo. \ref{algo:distributed_var_form} again after adding a control connection to each gate of the circuit representing $\Pi$ the distributed form, excluding the previously added non-local steps, to produce a circuit that achieves the controlled version of $\Pi$.

The remaining steps of $\alpha$-QPE are the two complications that arise which are discussed in Ref. \cite{wang2019accelerated} Section 2b. At each iteration of $\alpha$-QPE the Ansatz $\ket\psi = 1/\sqrt{2}(\ket{\phi} + \ket{-\phi})$ needs to be collapsed into either $\ket\phi$ or $\ket{-\phi}$.
In Ref. \cite{wang2019accelerated}, Wang et. al propose a statistical sampling method which one can apply a constant number of iterations in order to, with high confidence, both estimate the sign of  $\expval{P}{\psi}$ and ensure that  $|\expval{P}{\psi}| > \delta$. When this bound holds, then with high confidence, $\ket\psi$ can be efficiently collapsed to either one of $\ket\phi$ or $\ket{-\phi}$. Once this is performed, we apply the $\alpha$-QPE procedure as normal. If high confidence cannot be achieved, then instead of using the $\alpha$-QPE circuitry, statistical sampling continues. Statistical sampling in this setting implies repeatedly preparing $\ket\psi$, applying the single layer Pauli string $P$, in order to estimate $\expval{P}{\psi}$. When the bound does not hold, statistical sampling is performed until $\expval{P}{\psi}$ is estimated with sufficient precision in the normal VQE sense. We follow the method of Wang et. al, but use the modified $R(\lambda)$ circuit needed to prepare the Ansatz over a distributed quantum computer. We write this whole procedure in Algorithm $\ref{algo:dist_qpe}$.

\begin{definition}[Schedule]\label{def:sche}
  A schedule $S$ is a collection of $r$ lists where each element of a list contains the distribution of qubits on the $m$ QPUs. Each distribution is a list of qubit allocations on each QPU $q_i \in \{0, ..., Q_j\}$ where $Q_j$ is the number of qubits on QPU $j$. If the Ansatz is not allocated in a round $r' \in \{1,..., r\}$, it does not appear in the distribution list. The structure of a schedule is as follows:
  \begin{align*}
    S = \{\\ 
      & 1: [[q_{1}, ..., q_m]_1, ..., [q_1, ..., q_m]_{n_1}], 2: [[q_{1}, ..., q_m]_{n_1+1}, ..., [q_1, ..., q_m]_{n_2}], \\
      & ...,r: [[q_{1}, ..., q_m]_{n_{r-1}+1}, ..., [q_1, ..., q_m]_{n_r}]                                                    \\
    \}
  \end{align*}
  The subscripts on the qubit count lists represent the index of the Pauli being estimated.
\end{definition}

\begin{algo}[ht]
    \small
  \caption{Distributed $\alpha$-QPE}
  \textbf{Input:} 
  \begin{compactitem}
    \item $S$: A schedule (defined in Definition \ref{def:sche}) providing the qubit mapping of an Ansatz of size $n$ for $p$ Pauli strings on $m$ QPUs.
    \item $A = [a_1, ..., a_n]$: The vector constants for each Pauli string.
    \item $U = [U_1, ..., U_n]$: The circuits for $\alpha$-QPE
    \item $P = [P_1, ..., P_n]$: The Pauli operators associated with each $U_i$
  \end{compactitem}
  \textbf{Output:} The value of the expectation value estimations of the $p$ Paulis \\
  \textbf{Distributed $\alpha$-QPE}$(S, A, U, P)$:
  \begin{algorithmic}[1]
    \State $estimates \gets [\hspace{1mm}]$ 
    \For {$r \in S$}
    \State $roundOfEstimates \gets \textbf{RunAQPERound}(S(r), U)$ 
    \State $estimates$.\textbf{addAll}($roundOfEstimates$) \Comment{Order of estimates is fixed}
    \EndFor
    \State \textbf{return} $A \cdot estimates$ \Comment{Return the scalar product}
  \end{algorithmic}
  \textbf{RunAQPERound}$(S(r), U)$:
  \begin{algorithmic}[1]
    \State $estimates \gets Array(|S(r)|)$ \Comment{Initalize $|S(r)|$ length array  }
    \Parallel{$p \in S(r)$}
    \State $success \gets $ Bound $|\expval{P_{p}}{\psi(\lambda)}|$ away from $0$ and $1$ using \cite[Appendix C, Stage \RN{1}]{wang2019accelerated} 
    \If{$success$}
    \State Perform \cite[Appendix C, Stage \RN{2}]{wang2019accelerated} to collapse $\ket\psi$ to either $\ket\phi$ or $\ket{-\phi}$
    \State \parbox[t]{385pt}{$estimates[p] \gets $  Perform $\alpha$-QPE using distributed circuits with $c-U_p$ or $c-U_p^{\dagger}$ depending on collapsed $\ket\psi$}
    \Else
    \State \parbox[t]{385pt}{$estimates[p] \gets $ Estimate $\expval{P_{p}}{\psi(\lambda)}$ with statistical sampling using constant distribution circuits using the Ansatz distribution $p$}
    \EndIf
    \End
    \State \textbf{return} $estimates$
  \end{algorithmic}
  \label{algo:dist_qpe}
\end{algo}

\subsubsection{Distributed \texorpdfstring{$\alpha$}{a}-VQE}

To conclude the mapping of a localized, monolithic version of $\alpha$-VQE to the distributed version, we need to replace the $\alpha$-QPE subroutine with the distributed $\alpha$-QPE version from the previous section. For completeness we write distributed $\alpha$-VQE as an algorithm in Algorithm \ref{algo:dist_avqe}.

\begin{algo}[H]
    \small
    \caption{Distributed $\alpha$-VQE}
    \label{algo:dist_avqe}
    \textbf{Input}:
    \begin{compactitem}
    \item A list of QPU sizes $Q = [q_1, q_2, ..., q_m]$
    \item $H$ the Hamiltonian $H = A \cdot P = \sum_{i=1}^n a_i P_i$
    \item $R(\lambda)$ The Ansatz preparation circuit
  \end{compactitem}
  \textbf{Output}: An estimate for $\expval{H}{\psi(\lambda)}$, $\ket{\psi(\lambda)}$ the state prepared by circuit $R(\lambda)$.\\
  \textbf{Distributed $\alpha$-VQE}$(Q, H, R(\lambda))$
  \begin{algorithmic}[1]
      \State $q \gets$ Number of qubits needed for $R(\lambda)$
      \State $p \gets$ Number of Paulis for $H$
      \State $S, map \gets$ Ansatz schedule from an algorithm proposed in Section \ref{sec:distributing}
      \State $dR(\lambda) \gets \textbf{DistributedRemapper}(R(\lambda), map)$
      \State $dR(\lambda)^\dagger \gets \textbf{DistributedRemapper}(R(\lambda)^\dagger, map)$
      \State $d\Pi \gets \textbf{DistributedRemapper}(\Pi, map)$
      \State $c-\Pi \gets$ Add control connections to $d\Pi$ from pre-allocated $\alpha$-QPE qubit
      \State $c-d\Pi \gets \textbf{DistributedRemapper}(c-\Pi, map)$
      \For{$P_i \in P$}
          \State $dP_i \gets \textbf{DistributedRemapper}(P_i, map)$
          \State $c-dU_i \gets$ Combine distributed circuits $dR (c-d\Pi) dR^{\dagger} dP_i dR (c-d\Pi)  dP_i dR^{\dagger}$
      \EndFor
      \State $c-dU \gets [c-dU_1, ..., c-dU_n]$
      \State $dP \gets [dP_1, ..., dP_n]$
      \State $\expval{H}{\psi(\lambda)} \gets \textbf{Distributed } \alpha$-\textbf{QPE}$(S, A, c-dU, dP)$
      \State \textbf{return} $\expval{H}{\psi(\lambda)}$
  \end{algorithmic}
\end{algo}

\begin{subsection}{Analysis}
  In this section, we analyse the properties of the distributed quantum circuits in relation to the Ansatz size. First, we compare the duration of computation using three methods of performing the estimates of the expectation values: estimating in parallel, on one single QPU the size of the Ansatz, and using parallel and distributed computing. When running in parallel, one Pauli string is estimated per QPU. The limitation is that the Ansatz can be only as big as the smallest QPU, minus the qubit for $\alpha$-QPE. In the single QPU case, we assume the full Ansatz can fit on the QPU, and therefore no gates are distributed. Finally, in the distributed and parallel case, Pauli strings are estimated similarly to the parallel case, but multiple Ans\"{a}tze can be placed on a single QPU as well as split between multiple QPUs with distributed control gates. 
  
  To get an estimate for the number of gates used, we analye the pieces of the $U$ operator defined in the previous section. The reflection operator $\Pi$ has the equivalent cost, up to $2n$ single qubit gates to an $(n+1)$-qubit Toffoli gate \cite[Section \RN{2}.B]{wang2019accelerated}. Without ancilla qubits, currently the circuit depth to implement such a gate grows linearly $O(n)$ \cite{saeedi2013linear} with improved linear scaling with 1 ancilla qubit \cite{he2017decompositions}. When $\lceil\tfrac{n-2}{2}\rceil$ ancilla qubits are available, the depth can scale as $O(\log n)$ \cite{maslov2016advantages} to implement with $6n-6$ CNOT gates. The additional ancilla qubits to decrease the circuit depth could be considered in the Ansatz distribution phase from Subsection \ref{sec:distributing}, and we leave it to future work to analyse this change. Here we assume no additional ancilla qubits. For the Ansatz preparation $R(\lambda)$, in most of the applications to date, the circuit depth is $\Omega(n)$ \cite{babbush2018low}, meaning it has a tight upper and lower bound proportional to the number of qubits, which could be the most significant overhead in this process.
  
  We demonstrate the time trade-off. In Fig. \ref{fig:trade-off-1} we assume we have a QPU cluster with 5 QPUs each with 10 qubits. We determine a rough upper bound on the number of gates needed to perform distributed computing and summarize the time weight and gate quantity scaling in Table \ref{table:scaling-weighting}. In Fig. \ref{fig:num_qubits}, we show the maximum number of qubits that an Ansatz can be composed of using four different sized QPUs and with respect to the adding additional QPUs of the that size to the distributed system.
  
  \begin{table}[ht]
    \centering
    \begin{tabular}{|l|c|c|}
      \hline
      \textbf{Operation}      & \textbf{Execution time weight} & \textbf{ Quantity scaling } \\ \hline
      CNOT                    & 5                              & $O(n^4 \cdot \log n)$       \\ \hline
      Single qubit gate       & 1                              & $O(n^4 \cdot \log n)$       \\ \hline
      Measurements            & 2                              & $O(n^4 \cdot \log n)$       \\ \hline
      Entanglement generation & 8                              & $O(n^4 \cdot \log n)$       \\ \hline
      Classical communication & 2                              & $O(n^4 \cdot \log n)$       \\ \hline
      Output merging          & 3                              & $O(m)$                      \\ \hline
    \end{tabular}
    \caption{The time scaling of gates. $n$ represents the number of qubits in the Ansatz an $m$ the number of QPUs. The execution time weights are derived from Ref. \cite{michielsen2017benchmarking} for superconducting qubits. The quantity scalings are based on a Bravyi-Kitaev mapping \cite{seeley2012bravyi}.}
    \label{table:scaling-weighting}
  \end{table}
  
  \begin{figure}[H]
    \centering
\begin{tikzpicture}

\definecolor{color0}{rgb}{0.12156862745098,0.466666666666667,0.705882352941177}
\definecolor{color1}{rgb}{1,0.498039215686275,0.0549019607843137}
\definecolor{color2}{rgb}{0.172549019607843,0.627450980392157,0.172549019607843}

\begin{axis}[
legend cell align={left},
width=0.85\textwidth,
height=0.45\textwidth,
legend style={fill opacity=0.8, draw opacity=1, text opacity=1, at={(0.03,0.97)}, anchor=north west, draw=white!80!black},
tick align=outside,
tick pos=left,
grid=both,
x grid style={white!69.0196078431373!black},
xlabel={Ansatz size},
xmin=0.15, xmax=40.85,
xtick style={color=black},
y grid style={white!69.0196078431373!black},
ylabel={Weighted time },
ymin=-0.049999999860927, ymax=1.04999999999338,
ytick style={color=black}
]
\addplot [very thick, color0]
table {%
2 1.32450523815983e-10
2 1.32450523815983e-10
3 9.20628567471659e-10
4 3.61108980678824e-09
5 1.7649719823694e-07
6 1.64611017077353e-07
7 4.25643083671111e-07
8 1.0344086308173e-06
9 2.33414040053452e-06
10 4.61306342126603e-05
11 0.000108936398028272
12 0.000162665632414064
13 0.000269810336293574
14 0.000435625583607981
15 0.000673233299477425
16 0.00180420057498408
17 0.00264330194619639
18 0.00381301008670588
19 0.0060736399564462
20 0.00840635540799239
21 0.0114487914150632
22 0.0153661719470204
23 0.0203515553560946
24 0.0368707818065983
25 0.0477087348655738
26 0.0611023119727956
27 0.0775178796119449
28 0.0974840169283297
29 0.121596991660825
30 0.150526487212345
31 0.185021582449547
32 0.288671698278961
33 0.350289386800648
34 0.422580135704849
35 0.506989834818063
36 0.605108938377879
37 0.718682082477624
38 0.849618037912323
39 1
};
\addlegendentry{Distributed}
\addplot [very thick, color1]
table {%
2 4.4885668167384e-10
2 4.4885668167384e-10
3 8.10352992666723e-09
4 5.74536552542515e-08
5 2.54446518495216e-07
6 8.45842436246932e-07
7 2.3163985408929e-06
8 5.51555090440815e-06
9 1.18149466330808e-05
10 2.32979628348476e-05
11 4.29821941874038e-05
12 7.50757732599784e-05
13 0.000125268295489948
14 0.000201057046359391
15 0.000312109167087188
16 0.000470660343842829
17 0.000691950561861118
18 0.0009946974282826
19 0.00140160753487445
20 0.00193992630310408
21 0.0026420267286534
22 0.00354603741983663
23 0.00469651030409161
24 0.0061451283584098
25 0.00795145370297549
26 0.0101837163821701
27 0.0129196441432739
28 0.0162473335105048
29 0.020266162440342
30 0.0250877448332711
31 0.0308369271670686
32 0.0376528275074263
33 0.0456899171430339
34 0.0551191450841249
35 0.0661291056558952
36 0.0789272494110737
37 0.0937411375792248
38 0.110819740264047
39 0.130434778593979
};
\addlegendentry{One QPU}
\addplot [very thick, color2]
table {%
2 1.32450523815983e-10
2 1.32450523815983e-10
3 1.84125713494332e-09
4 1.4444359227153e-08
5 5.09904854660307e-08
6 1.88126876659832e-07
7 4.72936759634568e-07
8 1.12060935005207e-06
9 2.48002417556793e-06
};
\addlegendentry{Parallel}
\end{axis}

\end{tikzpicture}

    \caption{This plot is of a weighted time using the greedy distribution of the Ansatz for growing Ansatz sizes with 5 QPUs each with 10 qubits. The green line shows the timing for running 1 Ansatz per QPU. It cuts off at 9 qubits. The orange line is if all 50 qubits were on 1 QPU. The blue line is if we use a distributed Ansatz over the 5 QPUs.}
    \label{fig:trade-off-1}
  \end{figure}
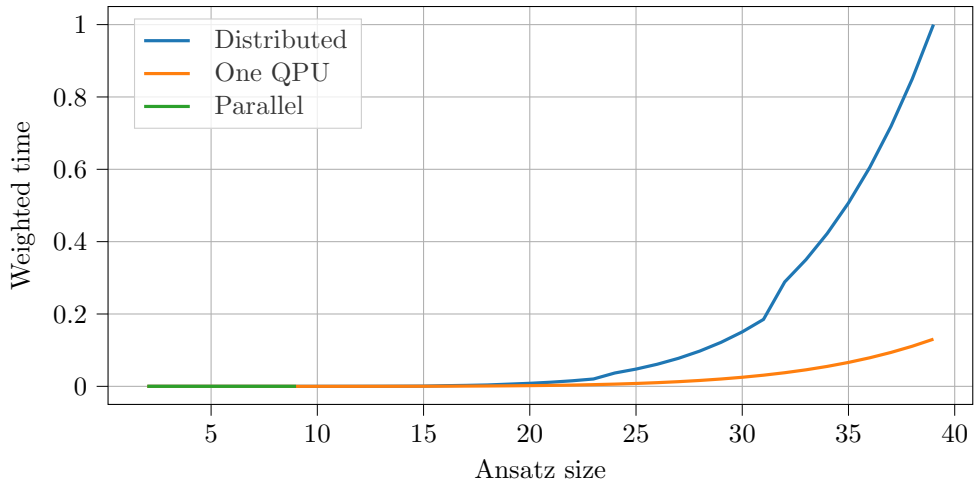
  
  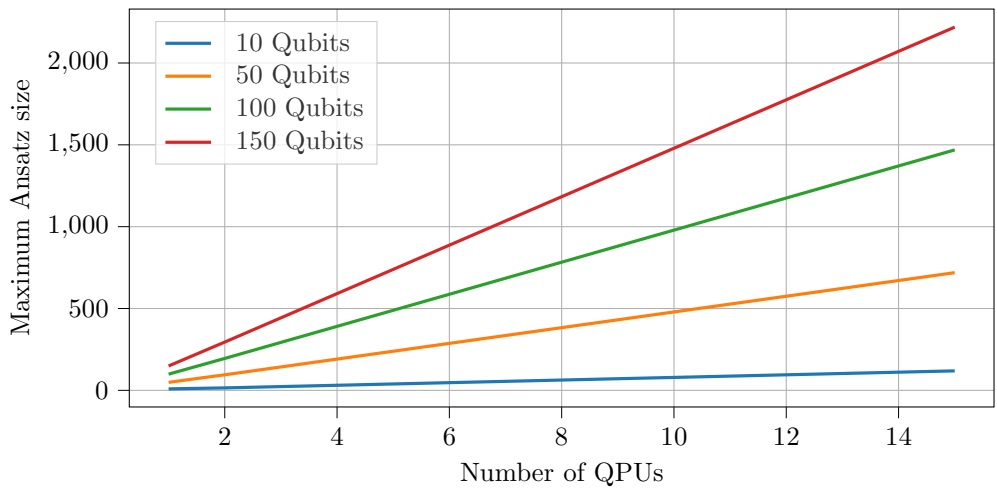
\begin{figure}[H]
    \centering
\begin{tikzpicture}[]

\definecolor{color0}{rgb}{0.12156862745098,0.466666666666667,0.705882352941177}
\definecolor{color1}{rgb}{1,0.498039215686275,0.0549019607843137}
\definecolor{color2}{rgb}{0.172549019607843,0.627450980392157,0.172549019607843}
\definecolor{color3}{rgb}{0.83921568627451,0.152941176470588,0.156862745098039}

\begin{axis}[
legend cell align={left},
width=0.85\textwidth,
height=0.45\textwidth,
legend style={fill opacity=0.8, draw opacity=1, text opacity=1, at={(0.03,0.97)}, anchor=north west, draw=white!80!black},
tick align=outside,
tick pos=left,
grid=both,
x grid style={white!69.0196078431373!black},
xlabel={Number of QPUs},
xmin=0.3, xmax=15.7,
xtick style={color=black},
y grid style={white!69.0196078431373!black},
ylabel={Maximum Ansatz size},
ymin=-101.5, ymax=2329.5,
ytick style={color=black}
]
\addplot [very thick, color0]
table {%
1 9
2 15
3 23
4 31
5 39
6 47
7 55
8 63
9 71
10 79
11 87
12 95
13 103
14 111
15 119
};
\addlegendentry{10 Qubits}
\addplot [very thick, color1]
table {%
1 49
2 95
3 143
4 191
5 239
6 287
7 335
8 383
9 431
10 479
11 527
12 575
13 623
14 671
15 719
};
\addlegendentry{50 Qubits}
\addplot [very thick, color2]
table {%
1 99
2 195
3 293
4 391
5 489
6 587
7 685
8 783
9 881
10 979
11 1077
12 1175
13 1273
14 1371
15 1469
};
\addlegendentry{100 Qubits}
\addplot [very thick, color3]
table {%
1 149
2 295
3 443
4 591
5 739
6 887
7 1035
8 1183
9 1331
10 1479
11 1627
12 1775
13 1923
14 2071
15 2219
};
\addlegendentry{150 Qubits}
\end{axis}

\end{tikzpicture}

    \caption{The maximum Ansatz size that would fit on a distributed system of QPUs. The maximum Ansatz size is given by $\sum_{i=1}^n q_i - 2n - 1$ with $n$ QPUs with $q_i > 2$ qubits on QPU $i$.}
    \label{fig:num_qubits}
  \end{figure}
\end{subsection}

\subsection{Applications for Quantum Chemistry}

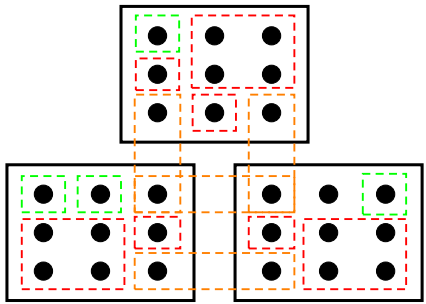
\begin{figure}[H]
  \centering
  \begin{tikzpicture}[scale=0.6, every node/.style={transform shape}, 
      qpu/.pic = {
        \draw[line width=.4mm] (.2, 0.85) rectangle (4.3, 3.85);
        \foreach \x in {1, 2.25, 3.5} \foreach \y in {0.85, 1.70, 2.55}
        \draw[fill] (\x, 0.65 + \y) circle (2mm);
    }]
    \draw (0,0) pic {qpu};
    \draw (5,0) pic {qpu};
    \draw (2.5, 3.5) pic {qpu};
        
    \draw [densely dashed, color=green, line width=0.25mm] (.525, 3.6) rectangle ++(0.95, -.8);    
    \draw [densely dashed, color=green, line width=0.25mm] (1.75, 3.6) rectangle ++(0.95, -.8);
    \draw [densely dashed, color=green, line width=0.25mm] (8.0, 3.65) rectangle ++(0.95, -.9);  
    \draw [densely dashed, color=green, line width=0.25mm] (3.025, 7.15) rectangle ++(0.95, -.8); 
        
    \draw [densely dashed, color=orange, line width=0.25mm] (3, 3.6) rectangle ++(3.5, -.8);
    \draw [densely dashed, color=orange, line width=0.25mm] (3, 1.9) rectangle ++(3.5, -.8);
    \draw [densely dashed, color=orange, line width=0.25mm] (3, 2.8) rectangle ++(1, 2.6);
    \draw [densely dashed, color=orange, line width=0.25mm] (5.5, 2.8) rectangle ++(1, 2.6);
        
    \draw [densely dashed, color=red, line width=0.25mm] (3.025, 6.2) rectangle ++(0.95, -.7); 
    \draw [densely dashed, color=red, line width=0.25mm] (4.27, 5.4) rectangle ++(0.95, -.8);
    \draw [densely dashed, color=red, line width=0.25mm] (3.0, 2.7) rectangle ++(1, -.7); 
    \draw [densely dashed, color=red, line width=0.25mm] (5.5, 2.7) rectangle ++(1, -.7); 
    \draw [densely dashed, color=red, line width=0.25mm] (.525, 2.65) rectangle ++(2.25, -1.55);    
    \draw [densely dashed, color=red, line width=0.25mm] (6.7, 2.65) rectangle ++(2.25, -1.55);
    \draw [densely dashed, color=red, line width=0.25mm] (4.25, 7.15) rectangle ++(2.25, -1.6);
  \end{tikzpicture}
  \caption{A distributed Ansatz of size 4 on three QPUs with 9 qubits. The green outlined qubits are reserved for running $\alpha$-QPE. The red outline qubits are for the Ans\"atze. The orange outlined are qubits reserved for entanglement between QPUs for non-local gates. One qubit is left idle.}
  \label{fig:chem_example}
\end{figure}

In this section, we take an example of a electronic molecular Hamiltonian for the chemical $H_2$. To estimate the Hamiltonian for this molecule with 2 electrons and 2 active orbitals, we require 4 qubits when using a Bravyi-Kitaev transformation. We can quickly obtain the Hamiltonian using the Pennylane Python library \cite{bergholm2018pennylane}. The Hamiltonian in this case, under the Bravyi-Kitaev transformation is of the form, 
\begin{align}
    H = \sum_{i=1}^{15} a_i P_i,
\end{align}
where we are concerned in the number of elements in the sum and less so about the constant factors and therefore to perform $\alpha$-VQE, we will need to estimate 15 Pauli strings. In this example, we will consider a distributed quantum system of 3 QPUs each containing 9 qubits. If we use these parameters as input to the algorithms in Section \ref{sec:ansatz_dist}, the output configuration would be the one depicted in Figure \ref{fig:chem_example}. In one round, 4  Ans\"{a}zte can fit across this distributed system, and so at least 4 rounds need to be executed. We can use the same allocation for the first 3 rounds and in the last round eliminate the distributed Ansatz in order to reduce the need for cross communication between QPUs. 

\begin{figure}[H]
  \centering
  \subfloat[Circuit representation of reflection $\Pi$.]{
    \begin{tikzpicture}[scale=1, every node/.style={transform shape} ]
      \node[scale=0.9] () at (0, 3.0) {
        \Qcircuit @C=1em @R=1em {
          \lstick{} & \gate{X} & \qw & \ctrl{1} & \qw & \gate{X} & \qw \\
          \lstick{} & \gate{X} & \qw & \ctrl{1} & \qw & \gate{X} & \qw\\
          \lstick{} & \gate{X} & \qw & \ctrl{1} & \qw & \gate{X} & \qw\\
          \lstick{} & \gate{X} & \gate{H} & \targ & \gate{H} &\gate{X} & \qw\\
        }
      };
      \node () at (0, 1.3) {\rotatebox{90}{$\,=$}};
      \node[scale=0.75] () at (0, 0){
        \Qcircuit @C=0.65em @R=0.6em  {
          \lstick{} & \gate{X} & \qw & \qw & \qw & \qw & \qw & \targ & \ctrl{3} & \targ & \ctrl{3} & \targ & \ctrl{3} & \targ & \ctrl{3} & \gate{X} & \qw \\
          \lstick{} & \gate{X} & \qw & \targ & \ctrl{2} & \targ & \ctrl{2} & \ctrl{-1} & \qw & \qw & \qw & \ctrl{-1} & \qw & \qw & \qw & \gate{X}  & \qw  \\
          \lstick{} & \gate{X} & \ctrl{1} & \ctrl{-1} & \qw & \ctrl{-1} & \qw & \qw & \qw & \ctrl{-2} & \qw & \qw & \qw & \ctrl{-2} & \qw & \gate{X} & \qw \\
          \lstick{} & \gate{X} & \gate{R(0,0,\frac{\pi}{2})} & \qw & \gate{R(0,\frac{-\pi}{2},0)} & \qw & \gate{R(0,0,\frac{\pi}{2})} & \qw & \gate{R(0,\frac{-\pi}{2},0)} & \qw & \gate{R(0,0,\frac{\pi}{2})} & \qw & \gate{R(0,\frac{-\pi}{2},0)} & \qw & \gate{R(0,0,\frac{\pi}{2})} & \gate{X} & \qw  \\
      }};
    \end{tikzpicture}
    }\\[10mm]
      \subfloat[Circuit representation of distributed $\Pi$. The square gates in the 4 qubit gates represent the cat-entangler/disentager sequence.]{
    \begin{tikzpicture}[scale=1, every node/.style={transform shape} ]
      \node[scale=0.725] () at (0, 0){
        \Qcircuit @C=0.6em @R=1em  {
          \lstick{} & \gate{X} & \qw & \qw & \qw & \qw & \qw & \targ & \ctrl{1} & \targ & \ctrl{1} & \targ & \ctrl{1} & \targ & \ctrl{1} & \gate{X} & \qw \\
          \lstick{\ket{0}} & \qw & \qw & \qw & \qw & \qw  & \qw & \gate{} \qwx[-1] & \gate{}\qwx[3] & \gate{} \qwx[-1] & \gate{} \qwx[3]  & \gate{}\qwx[-1] & \gate{} \qwx[3] & \gate{}\qwx[-1] \qwx[3] & \gate{}\qwx[3] & \qw & \qw   \\
          \lstick{\ket{0}} & \qw & \qw & \gate{} & \gate{} & \gate{}  & \gate{} & \gate{} \qwx[-1] & \qw & \qw & \qw  & \gate{} \qwx[-1] & \qw  & \qw & \qw & \qw & \qw \\
          \lstick{} & \gate{X} & \qw & \targ \qwx[-1] \qwx[1] & \ctrl{1} \qwx[-1] & \targ \qwx[-1] & \ctrl{1}\qwx[-1] & \ctrl{-1} & \qw & \qw & \qw & \ctrl{-1} & \qw & \qw & \qw & \gate{X}  & \qw  \\
          \lstick{\ket{0}} & \qw & \qw & \gate{} & \gate{} \qwx[2] & \gate{} \qwx[-1] & \gate{} \qwx[2] & \qw & \gate{} \qwx[2] & \gate{}\qwx[-3] & \gate{} \qwx[2] & \qw & \gate{} \qwx[2] & \gate{}  & \gate{} \qwx[2] & \qw & \qw \\
          \lstick{} & \gate{X} & \ctrl{1} & \ctrl{-1} & \qw & \ctrl{-1} & \qw & \qw & \qw & \ctrl{-1} & \qw & \qw & \qw & \ctrl{-1} & \qw & \gate{X} & \qw \\
          \lstick{} & \gate{X} & \gate{R(0,0,\frac{\pi}{2})} & \qw & \gate{R(0,\frac{-\pi}{2},0)} & \qw & \gate{R(0,0,\frac{\pi}{2})} & \qw & \gate{R(0,\frac{-\pi}{2},0)} & \qw & \gate{R(0,0,\frac{\pi}{2})} & \qw & \gate{R(0,\frac{-\pi}{2},0)} & \qw & \gate{R(0,0,\frac{\pi}{2})} & \gate{X} & \qw  \\
      }};
      \draw[color=red!100, dashed, line width=0.15mm] (-7.45, 1.9) rectangle (7.1, 0.95);
      \draw[color=red!100, dashed, line width=0.15mm] (-7.45, 0.875) rectangle (7.1, -0.1);
      \draw[color=red!100, dashed, line width=0.15mm] (-7.45, -.2) rectangle (7.1, -1.9);
    \end{tikzpicture}
  }
  \caption{Distributed circuit mapping for reflection $\Pi$.}
  \label{fig:reflection}
\end{figure}
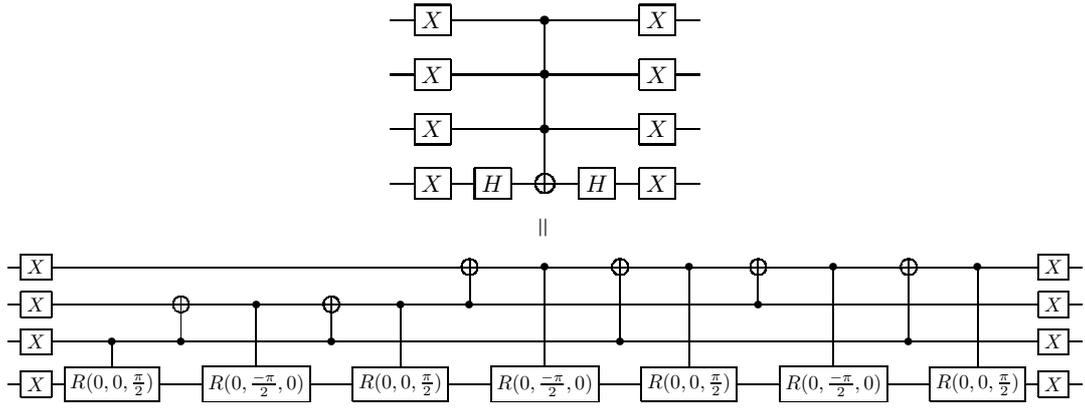
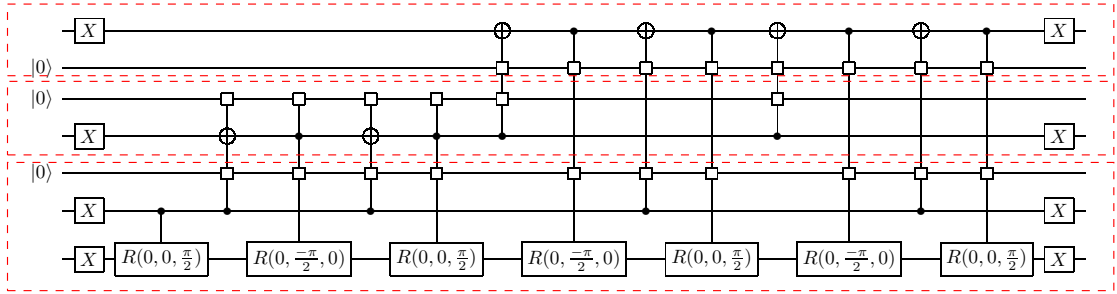

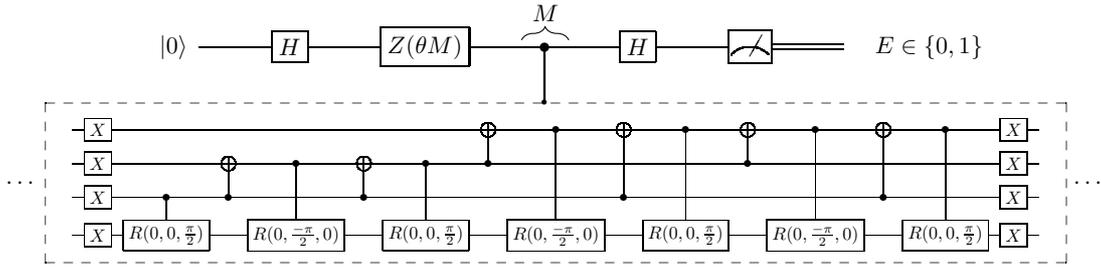
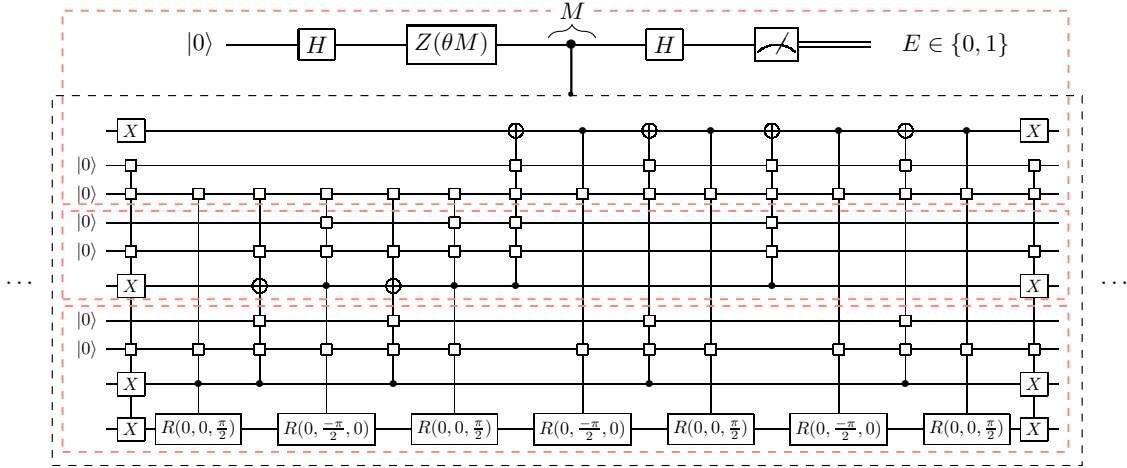
\begin{figure}[H]
    \centering
    \subfloat[To run $\alpha$-QPE, one needs to perform a controlled $U$ operation $M$ times, where $U=R\Pi R^{\dagger}PR\Pi R^{\dagger}P$. The control portion to consider is $c-\Pi$. We depict the $c-\Pi$ part, where the other parts of $U$ are applied before and after what is depicted, which do not need to be controlled.]{
    \begin{tikzpicture}[scale=0.90, every node/.style={transform shape} ]
      \node at (-0.5, 2) {
          \Qcircuit @C=3em @R=1em  {
              \lstick{\ket 0} & \gate{H} & \gate{Z(\theta M)} & \ctrl{0} & \gate{H} & \meter & \cw  
          }
      };
      \node[scale=2] at (.33 - 0.5, 1.17) {$\cdot$};
      \node[scale=1] at (.33 - 0.48, 2.5) {$M$};
      \node at (5.45, 2.) {$E\in\{0, 1\}$};
      \node at (-7.8, 0) {\dots};
      \node at (7.8, 0) {\dots};
      \draw [decorate, decoration={brace,amplitude=5pt,}] (0  - 0.5, 2.10) -- (.69 - 0.5, 2.1);
      \draw[line width=0.3mm] (0.33  - 0.495, 2) -- (.33 - 0.495, 1.19);
      \node[scale=0.75] () at (0, 0){
        \Qcircuit @C=0.65em @R=0.6em  {
          \lstick{} & \gate{X} & \qw & \qw & \qw & \qw & \qw & \targ & \ctrl{3} & \targ & \ctrl{3} & \targ & \ctrl{3} & \targ & \ctrl{3} & \gate{X} & \qw \gategroup{1}{1}{4}{17}{3em}{--}\\
          \lstick{} & \gate{X} & \qw & \targ & \ctrl{2} & \targ & \ctrl{2} & \ctrl{-1} & \qw & \qw & \qw & \ctrl{-1} & \qw & \qw & \qw & \gate{X}  & \qw  \\
          \lstick{} & \gate{X} & \ctrl{1} & \ctrl{-1} & \qw & \ctrl{-1} & \qw & \qw & \qw & \ctrl{-2} & \qw & \qw & \qw & \ctrl{-2} & \qw & \gate{X} & \qw \\
          \lstick{} & \gate{X} & \gate{R(0,0,\frac{\pi}{2})} & \qw & \gate{R(0,\frac{-\pi}{2},0)} & \qw & \gate{R(0,0,\frac{\pi}{2})} & \qw & \gate{R(0,\frac{-\pi}{2},0)} & \qw & \gate{R(0,0,\frac{\pi}{2})} & \qw & \gate{R(0,\frac{-\pi}{2},0)} & \qw & \gate{R(0,0,\frac{\pi}{2})} & \gate{X} & \qw  \\
      }};
    \end{tikzpicture}
    }\\[10mm]
    \subfloat[Distributed $c-\Pi$. The square gates in the 4 qubit gates represent the cat-entangler/disentager sequence. ]{
    \begin{tikzpicture}[scale=0.90, every node/.style={transform shape} ]
        \node[scale=1] at (-0.5, 3.5) {
        \Qcircuit @C=3em @R=1em  {
         \lstick{\ket 0} & \gate{H} & \gate{Z(\theta M)} & \ctrl{0} & \gate{H} & \meter & \cw 
         }
        };
      \node[scale=0.75] () at (0, 0){
        \Qcircuit @C=0.6em @R=1em  {
          \lstick{} & \gate{X} & \qw & \qw & \qw & \qw & \qw & \targ & \ctrl{2} & \targ \qwx[1] & \ctrl{2} & \targ & \ctrl{2} & \targ & \ctrl{2} & \gate{X} & \qw \\
          \lstick{\ket{0}} & \gate{}\qwx[1] & \qw & \qw & \qw & \qw  & \qw & \gate{} \qwx[-1] & \qw & \gate{} & \qw  & \gate{}\qwx[-1] & \qw & \gate{}\qwx[-1]  & \qw & \gate{}\qwx[1] & \qw   \\
          \lstick{\ket{0}} & \gate{} \qwx[2] & \gate{}\qwx[5] & \gate{} \qwx[2] & \gate{} \qwx[1] & \gate{} \qwx[2]  & \gate{} \qwx[1] & \gate{} \qwx[-1] & \gate{}\qwx[5] & \gate{} \qwx[-1] & \gate{} \qwx[5]  & \gate{}\qwx[-1] & \gate{} \qwx[5] & \gate{}\qwx[-1] \qwx[4] & \gate{}\qwx[5] & \gate{}\qwx[2] & \qw   \\
          \lstick{\ket{0}} & \qw & \qw & \qw & \gate{} \qwx[1] & \qw  & \gate{} \qwx[1] & \gate{} \qwx[-1] & \qw & \qw & \qw  & \gate{} \qwx[-1] & \qw  & \qw & \qw & \qw & \qw \\
          \lstick{\ket{0}} & \gate{}\qwx[1] & \qw & \gate{} & \gate{} & \gate{}  & \gate{} & \gate{} \qwx[-1] & \qw & \qw & \qw  & \gate{} \qwx[-1] & \qw  & \qw & \qw & \gate{}\qwx[1] & \qw \\
          \lstick{} & \gate{X} \qwx[2] & \qw & \targ \qwx[-1] \qwx[1] & \ctrl{2} \qwx[-1] & \targ \qwx[-1] & \ctrl{2}\qwx[-1] & \ctrl{-1} & \qw & \qw & \qw & \ctrl{-1} & \qw & \qw & \qw & \gate{X} \qwx[2] & \qw  \\
          \lstick{\ket{0}} & \qw & \qw & \gate{} & \qw & \gate{} \qwx[-1] & \qw & \qw & \qw & \gate{}\qwx[-4] & \qw & \qw & \qw & \gate{}  & \qw & \qw & \qw \\
          \lstick{\ket{0}} & \gate{} \qwx[1] & \gate{}\qwx[1] & \gate{} \qwx[-1] & \gate{} \qwx[2] & \gate{} \qwx[-1] & \gate{} \qwx[2] & \qw & \gate{} \qwx[2] & \gate{}\qwx[-1] & \gate{} \qwx[2] & \qw & \gate{} \qwx[2] & \gate{} \qwx[-1]  & \gate{} \qwx[2] & \gate{} \qwx[1] & \qw \\
          \lstick{} & \gate{X}\qwx[1] & \ctrl{1} & \ctrl{-1} & \qw & \ctrl{-1} & \qw & \qw & \qw & \ctrl{-1} & \qw & \qw & \qw & \ctrl{-1} & \qw & \gate{X}\qwx[1] & \qw \\
          \lstick{} & \gate{X} & \gate{R(0,0,\frac{\pi}{2})} & \qw & \gate{R(0,\frac{-\pi}{2},0)} & \qw & \gate{R(0,0,\frac{\pi}{2})} & \qw & \gate{R(0,\frac{-\pi}{2},0)} & \qw & \gate{R(0,0,\frac{\pi}{2})} & \qw & \gate{R(0,\frac{-\pi}{2},0)} & \qw & \gate{R(0,0,\frac{\pi}{2})} & \gate{X} & \qw  \\
      }};
      \node at (5.45, 3.5) {$E\in\{0, 1\}$};
      \node at (-8.2, 0) {\dots};
      \node at (7.8, 0) {\dots};
      \draw[line width=0.3mm] (0.33  - 0.495, 3.5) -- (.33 - 0.495, 2.75);
      \draw[color=red!50, dashed, line width=0.25mm] (-7.6, 4.0) rectangle (7.1, 1.15);
      \draw[color=red!50, dashed, line width=0.25mm] (-7.6, 1.05) rectangle (7.1, -0.25);
      \draw[color=red!50, dashed, line width=0.25mm] (-7.6, -.35) rectangle (7.1, -2.5);
      \draw[dashed, line width=0.15mm] (-7.75, 2.75) rectangle (7.3, -2.7);
      \node[scale=1, fill=white] at (.33 - 0.48, 4.0) {$M$};
      \node[scale=2] at (.33 - 0.5, 2.75) {$\cdot$};
      \draw [decorate, decoration={brace,amplitude=5pt,}] (0  - 0.5, 3.6) -- (.69 - 0.5, 3.6);
    \end{tikzpicture}}
    \caption{Distributed circuit mapping for $c-\Pi$.}
    \label{fig:aqpe_circuit}
\end{figure}

For the Ansatz preparation, we use the circuit $R(\lambda)$ depicted in Fig. \ref{fig:ansatz_prep} (a). From the 4 Ans\"{a}tze, three of them will be able to run the $\alpha$-QPE step without distribution of the Ansatz. The fourth Ansatz is on the other hand distributed and will need to use the circuit in Fig. \ref{fig:ansatz_prep} (b) for preparation. For simplicity, we include arbitrary qubit rotations which are represented by the $R(\lambda_1,\lambda_2,\lambda_3)$ gates, where $\lambda_i \in [-\tfrac{\pi}{2},\tfrac{\pi}{2}]$ for $i \in \{1,2,3\}$. Next we need to perform the reflection operation $\Pi$ described in Section \ref{sec:aqpe}, whose circuit is shown in Figure \ref{fig:reflection} (a). An equivalent circuit is also shown which decomposes the 4-qubit Toffoli gate into a series controlled and single qubit gates. We again need a distributed version of the reflection operation to support the Ansatz which is distributed. We show this circuit in \ref{fig:reflection} (b). Here we introduce gates for the cat-entangler and cat-disentagler sequences. Here, 4 qubits are allocated for performing the non-local gates. Now, for running $\alpha$-QPE, we need a circuit for $c-\Pi$, which is the control part of $c-U$. Here is where it is critical to have 2 entanglement qubits for each splitting of the Ansatz on each QPU since, as seen in Figure \ref{fig:aqpe_circuit}, there are control-control gates that occur across QPUs. With this collection of gates, we can run $\alpha$-QPE and therefore using the algorithm in Section $\ref{sec:avqe}$ run $\alpha$-VQE.  

\begin{figure}[H]
  \centering
  \subfloat[Circuit for $R(\lambda), \lambda_i \in \text{[}-\pi, \pi\text{]}^3, i \in\{1,...,4\}$.]{
    \begin{tikzpicture}[scale=0.9,  every node/.style={transform shape}]
    \node at (0, 0) {\Qcircuit @C=1.9em @R=0.8em {
      \lstick{\ket{0}} & \gate{X} &\gate{R(\lambda_1)} &  \qw & \targ & \qw  & \qw\\
      \lstick{\ket{0}} & \gate{X} & \gate{R(\lambda_2)} &  \qw & \qw & \targ & \qw \\
      \lstick{\ket{0}} & \qw & \gate{R(\lambda_3)} &  \ctrl{1} & \ctrl{-2} & \qw &  \qw  \\
      \lstick{\ket{0}} & \qw & \gate{R(\lambda_4)} &  \targ & \qw & \ctrl{-2} &  \qw  \\
    }};
    \end{tikzpicture}
    } \\[10mm]
  \subfloat[Circuit for a distributed $R(\lambda)$. The red dashed lines represent the individual QPUs.]{
    \begin{tikzpicture}[scale=0.8,  every node/.style={transform shape}]
      \node at (0,0) {
        \Qcircuit @C=0.7em @R=1em {
          \lstick{\ket{0}} & \gate{X} & \gate{R(\lambda_1)} & \qw & \qw & \qw & \qw & \targ & \qw & \qw & \qw & \qw& \qw & \qw & \qw & \qw& \qw & \qw & \qw & \qw& \qw\\
          \lstick{\ket{0}} & \qw & \ctrl{0} & \qw & \qw & \qw & \targ & \ctrl{-1} & \gate{H} &\meter &\cctrl{4} & \targ& \qw& \qw& \qw& \qw& \qw& \qw& \qw& \qw & \qw\\ 
          \lstick{\ket{0}} & \gate{X} & \gate{R(\lambda_2)} & \qw & \qw & \qw & \qw & \qw & \qw & \qw & \qw & \qw & \qw & \qw & \qw & \targ & \qw & \qw & \qw & \qw & \qw \\
          \lstick{\ket{0}} & \qw& \qw & \qw & \qw & \qw & \qw & \qw & \qw & \qw & \qw & \ctrl{0} & \qw & \qw &\targ &\ctrl{-1} & \gate{H} & \meter & \cctrl{3} & \targ & \qw \\
          \lstick{\ket{0}} & \qw& \ctrl{0} & \qw & \targ & \meter & \cctrl{-3} & \targ & \qw & \qw & \qw & \ctrl{0} &  \targ & \meter & \cctrl{-1} & \qw & \qw& \qw & \qw & \qw& \qw \\
          \lstick{\ket{0}} & \qw & \gate{R(\lambda_3)} & \ctrl{1} & \ctrl{-1} & \qw & \qw & \qw & \qw  & \qw  & \gate{Z} & \qw & \qw & \qw & \qw & \qw & \qw& \qw & \qw& \qw& \qw \\
          \lstick{\ket{0}} & \qw & \gate{R(\lambda_4)} & \targ & \qw & \qw& \qw & \qw & \qw & \qw & \qw & \qw & \ctrl{-2} & \qw & \qw & \qw & \qw  & \qw & \gate{Z} & \qw & \qw\\
      }};
      \draw[color=red!50, dashed, line width=0.25mm] (-7.6, 3.1) rectangle (7, 1.4);
      \draw[color=red!50, dashed, line width=0.25mm] (-7.6, 1.25) rectangle (7, -0.3);
      \draw[color=red!50, dashed, line width=0.25mm] (-7.6, -.45) rectangle (7, -3.2);
      \draw[snake=coil,segment aspect=0,segment length=4pt] (1.1,-0.9) -- (1.1, 0.06);
      \draw[snake=coil,segment aspect=0,segment length=4pt] (-5.3,-0.88) -- (-5.3, 0.62);
      \draw[snake=coil,segment aspect=0,segment length=4pt] (-5.3,1.2) -- (-5.3, 1.8);
    \end{tikzpicture}
  }
  \caption{Distributed circuit mapping for $R(\lambda)$.}
  \label{fig:ansatz_prep}
\end{figure}
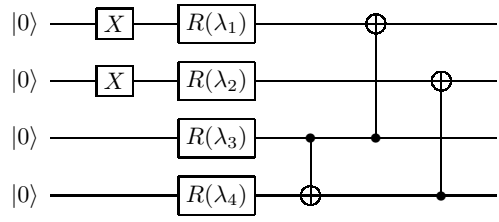
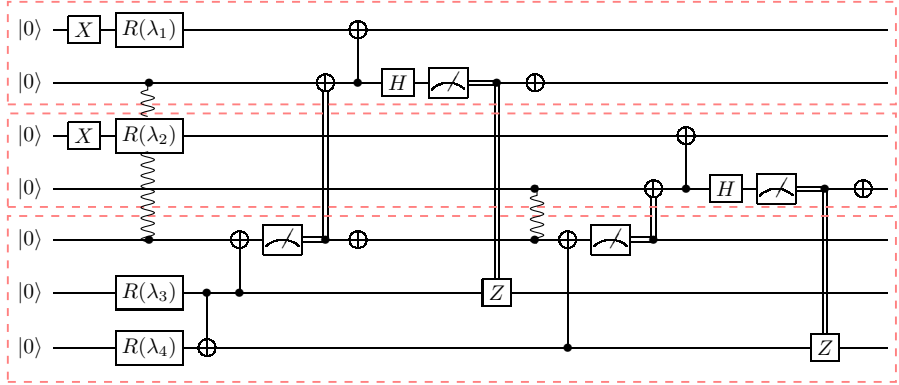

\clearpage
\section{Networked Control Systems for Distributed  QC}\label{sec:network_control}

Because it will be difficult in the near future to construct large, monolithic quantum computers, it will be a viable option to instead connect smaller quantum computers together using a network in a distributed manner. One can therefore consider networked control systems (NCSs) to manage the distribution of resources for running quantum algorithms. Such a system could allow for more flexibility regarding hardware configurations and the ability to dynamically add more devices while minimizing integration overheads. A NCS is a network of devices connected together using the network in order to perform a specific mutual task orchestrated by a control system \cite{hespanha2007survey, yang2006networked}. Among the other thing, NCSs are used to perform distributed or parallel computing, controlling a fleet of robots or drones, or smart grid systems deployed in modern cities \cite{mahmoud2019networked}.

Networked control systems can have various architectures for the control system part. These systems can either have a centralised controller where communications amongst the nodes are restricted to local area network (LAN) or a decentralised controller system that is connected via an internet or wide area network (WAN). These two scenarios resemble how distributed quantum computers could potentially be networked. In the first case, one can consider a single owner of multiple quantum devices where all of the quantum devices are located in the same room or building, specifically, the network owner would know the network topology and information about hardware in the network. In the second setting, multiple quantum computers located possibly far apart potentially connected by multi-hop connections where the owner of the hardware between the hops is possibly different. Here, more advanced protocols that consider security and robustness will be needed potentially leading to a fully fledged quantum Internet.

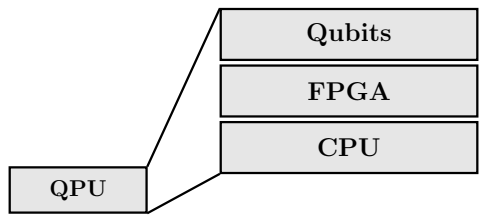
\begin{figure}[ht]
    \centering
    \begin{tikzpicture}[scale=0.75, every node/.style={transform shape}]
    \node[fill=gray!20, draw=black, minimum width=3cm, minimum height=1.0cm, line width=.3mm, scale=.8] (Q1) at (0,0.25) {\Large{\textbf{QPU}}};
    
    \node[scale=0.9, fill=gray!20, draw=black, minimum width=5cm, minimum height=1cm, line width=.3mm] (CPU) at (4.75, 1) {\Large{\textbf{CPU}}};
    \node[scale=0.9, fill=gray!20, draw=black, minimum width=5cm, minimum height=1cm, line width=.3mm] (FPGA) at (4.75, 2) {\Large{\textbf{FPGA}}};
    \node[scale=0.9 , fill=gray!20, draw=black, minimum width=5cm, minimum height=1cm, line width=.3mm] (qubits) at (4.75, 3) {\Large{\textbf{Qubits}}};
    
    \draw[-, line width=0.3mm] (Q1.north east) -- (qubits.north west);
    \draw[-, line width=0.3mm] (Q1.south east) -- (CPU.south west); 
    
    \end{tikzpicture}
    \caption{Internal layering of a QPU. We assume there is a layered architecture. The CPU instructs the FPGA which in turn controls and measures the qubits. The CPU also interfaces with the network.}
    \label{fig:qpu_internals}
\end{figure}

In order to use a network of distributed quantum computers efficiently, it is important that one develops robust communication protocols such that communication and control between the quantum devices in the network is efficient and reliable. In this section, we consider quantum systems with classical control and a separated quantum processing units. We consider a QPU to be a combination of a three layered system depicted in Fig. \ref{fig:qpu_internals}. The QPU in this case is a layered system with inputs and outputs to a communication network through a classical computer, or a CPU. The CPU interfaces with the network as well as controls the FPGA based on the control instructions from the network which in turn controls the qubits to perform quantum operations on the qubit layer. Qubit measurements and other classical messages are transmitted back to the network via a reversed path.

We consider the two different network configurations described and get into more detail about how these systems could be implemented in practice. We list the communication requirements needed to perform distributed quantum computations. We explore some available protocols to achieve these requirements under two scenarios. In the first one, there is a centralised controller of the system and communication to devices is classical information and quantum entanglement can be sent directly to other quantum processors without routing. The second case is when control over the network is not centralised, but has a single user. We then propose a control system using Deltaflow.OS to control system to orchestrate distributed quantum computing. 

\begin{figure}[ht]
    \centering
    \begin{tikzpicture}[scale=0.55, every node/.style={transform shape}]
        \node[fill=gray!20, draw=black, minimum width=4cm, minimum height=1.25cm, line width=.3mm] (Q1) at (0,0) {\Large{\textbf{QPU $1$}}};
        \node[fill=gray!20, draw=black, minimum width=4cm, minimum height=1.25cm, line width=.3mm] (Q2) at (7,0) {\Large{\textbf{QPU $2$}}};
        \node (dots1) at (11, 0)  {\Large{$\cdot\cdot\cdot$}};
        \node[fill=gray!20, draw=black, minimum width=4cm, minimum height=1.25cm, line width=.3mm] (Q3) at (15,0) {\Large{\textbf{QPU $n$}}};
    
        \node[fill=gray!60, draw=black, minimum width=19cm, minimum height=0.75cm, line width=.3mm] (net) at (7.5,-3) {\Large{\textbf{Classical Network}}};
        
        \node[fill=gray!20, draw=black, minimum width=3cm, minimum height=1.05cm, line width=.3mm, below = of net] (ctrl)  {\Large{\textbf{Controller}}};
        
        \draw[{latex[width=1.75mm]}-{latex[width=1.75mm]}, line width=0.4mm]  (Q1.south) -- (Q1 |- net.north);
        \draw[{latex[width=1.75mm]}-{latex[width=1.75mm]}, line width=0.4mm]  (Q2.south) -- (Q2 |- net.north);
        \draw[{latex[width=1.75mm]}-{latex[width=1.75mm]}, line width=0.4mm]  (Q3.south) -- (Q3 |- net.north);
        
        \draw[{latex[width=1.75mm]}-{latex[width=1.75mm]}, line width=0.4mm]  (ctrl.north) -- (ctrl |- net.south);
        
        \draw[snake=coil, segment aspect=0, segment length=4pt, line width=0.2mm]  ([yshift=2mm]Q1.east) -- ([yshift=2mm]Q2.west);
        \draw[snake=coil, segment aspect=0, segment length=4pt, line width=0.2mm]  ([yshift=2mm]Q2.east) -- ([yshift=2mm]dots1.west);
        \draw[snake=coil, segment aspect=0, segment length=4pt, line width=0.2mm]  ([yshift=2mm]dots1.east) -- ([yshift=2mm]Q3.west);
        
        \draw[{latex[width=1.75mm]}-{latex[width=1.75mm]}, line width=0.4mm]  ([yshift=-2mm]Q1.east) -- ([yshift=-2mm]Q2.west);
        \draw[{latex[width=1.75mm]}-, line width=0.4mm]  ([yshift=-2mm]Q2.east) -- ([yshift=-2mm]dots1.west);
        \draw[-{latex[width=1.75mm]}, line width=0.4mm]  ([yshift=-2mm]dots1.east) -- ([yshift=-2mm]Q3.west);
        
        \node[scale=1.1] at (3.5, -2.10) {Controller Communication};
        \node[scale=1.1] at (11.5, -2.10) {Controller Communication};
        \node[scale=1.1] at (11, .575) {Entanglement};
        \node[scale=1.1] at (3.5, .575) {Entanglement};
        \node[scale=1.1] at (11, -.65) {Low-Latency};
        \node[scale=1.1] at (3.5, -.65) {Low-Latency};
    \end{tikzpicture}
    
    \caption{A networked control system with a centralised controller. The triangular arrowheads represent classical connections and the diamond shaped arrow heads represent quantum connections used for establishing EPR pairs. Here we assume the network between the QPUs is completely connected in terms of quantum and classical connections, that is, each QPU has the same connections with any other QPU. In the completely connected network, this network is single use for transmitting with low latency. Moreover, each QPU is connected to a common bus line that handles all the latency-tolerant message exchanges among the nodes.}
    \label{fig:single_control}
\end{figure}
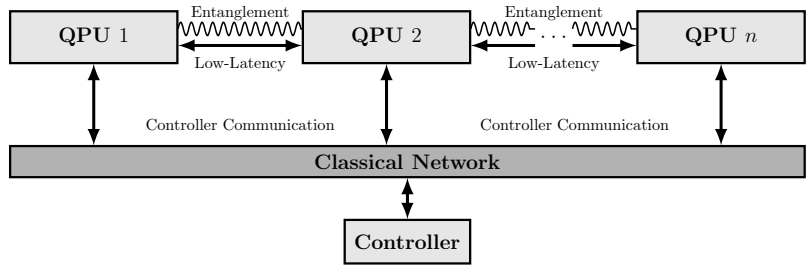

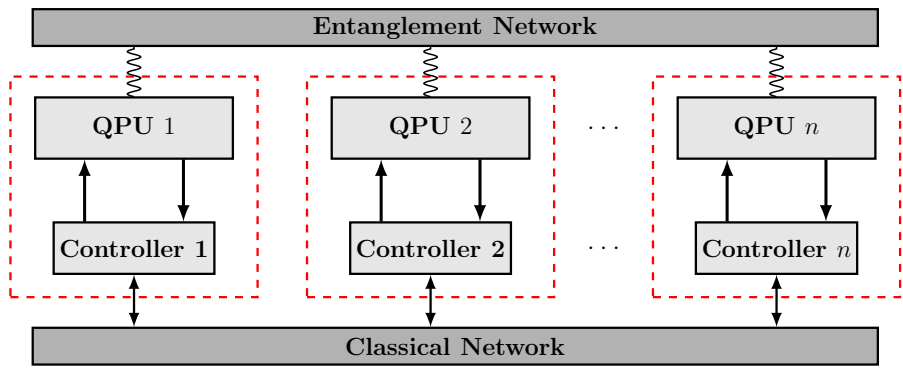
\begin{figure}
    \centering
    \begin{tikzpicture}[scale=0.65, every node/.style={transform shape}]
    \node[fill=gray!20, draw=black, minimum width=4cm, minimum height=1.25cm, line width=.3mm] (Q1) at (0,.20) {\Large{\textbf{QPU $1$}}};
    \node[fill=gray!20, draw=black, minimum width=4cm, minimum height=1.25cm, line width=.3mm] (Q2) at (6,.20) {\Large{\textbf{QPU $2$}}};
    \node[fill=gray!20, draw=black, minimum width=4cm, minimum height=1.25cm, line width=.3mm] (Q3) at (13,.20) {\Large{\textbf{QPU $n$}}};
    
    \node[fill=gray!60, draw=black, minimum width=17.1cm, minimum height=0.75cm, line width=.3mm] (net) at (6.5,-4.25) {\Large{\textbf{Classical Network}}};
    \node[fill=gray!60, draw=black, minimum width=17.1cm, minimum height=0.75cm, line width=.3mm] (ent_net) at (6.5, 2.25) {\Large{\textbf{Entanglement Network}}};
    
    \node[fill=gray!20, draw=black, minimum width=3.2cm, minimum height=1.05cm, line width=.3mm] (ctrl1) at (0.0,-2.25) {\Large{\textbf{Controller 1}}};
    \node[fill=gray!20, draw=black, minimum width=3.2cm, minimum height=1.05cm, line width=.3mm] (ctrl2) at (6,-2.25) {\Large{\textbf{Controller 2}}};
    
    \node at (9.5, .20) {\Large{$\cdot\cdot\cdot$}};
    \node at (9.5, -2.25) {\Large{$\cdot\cdot\cdot$}};
    
    \node[fill=gray!20, draw=black, minimum width=3.2cm, minimum height=1.05cm, line width=.3mm] (ctrl3) at (13, -2.25) {\Large{\textbf{Controller $n$}}};
    
    \draw[-{latex[width=1.75mm]}, line width=0.4mm]  ([xshift=1cm]Q1.south) -- ([xshift=1cm]Q1 |- ctrl1.north);
    \draw[{latex[width=1.75mm]}-, line width=0.4mm]  ([xshift=-1cm]Q1.south) -- ([xshift=-1cm]Q1 |- ctrl1.north);
    
    \draw[-{latex[width=1.75mm]}, line width=0.4mm]  ([xshift=1cm]Q2.south) -- ([xshift=1cm]Q2 |- ctrl2.north);
    \draw[{latex[width=1.75mm]}-, line width=0.4mm]  ([xshift=-1cm]Q2.south) -- ([xshift=-1cm]Q2 |- ctrl2.north);
    
    \draw[-{latex[width=1.75mm]}, line width=0.4mm]  ([xshift=1cm]Q3.south) -- ([xshift=1cm]Q3 |- ctrl3.north);
    \draw[{latex[width=1.75mm]}-, line width=0.4mm]  ([xshift=-1cm]Q3.south) -- ([xshift=-1cm]Q3 |- ctrl3.north);

    \node[dashed, line width=.3mm, minimum width=5cm, minimum height=4.5cm, draw=red] at (0, -1) {};
    \node[dashed, line width=.3mm, minimum width=5cm, minimum height=4.5cm, draw=red] at (6, -1) {};
    \node[dashed, line width=.3mm, minimum width=5cm, minimum height=4.5cm, draw=red] at (13, -1) {};

    \draw[{latex[width=1.25mm]}-{latex[width=1.25mm]}, line width=0.3mm] (ctrl1) -- (ctrl1 |- net.north);
    \draw[{latex[width=1.25mm]}-{latex[width=1.25mm]}, line width=0.3mm] (ctrl2) -- (ctrl2 |- net.north);
    \draw[{latex[width=1.25mm]}-{latex[width=1.25mm]}, line width=0.3mm] (ctrl3) -- (ctrl3 |- net.north);
    
    \draw[snake=coil, segment aspect=0, segment length=4pt, line width=0.2mm] (Q1) -- (Q1 |- ent_net.south);
    \draw[snake=coil, segment aspect=0, segment length=4pt, line width=0.2mm] (Q2) -- (Q2 |- ent_net.south);
    \draw[snake=coil, segment aspect=0, segment length=4pt, line width=0.2mm] (Q3) -- (Q3 |- ent_net.south);
    
    \end{tikzpicture}
    \caption{An inter-networked distributed quantum computer with decentralized controllers. Here, there are independent controllers that control their respective quantum processing systems. Entanglement is generated with an entanglement network requiring possibly multi-hop entanglement routing. The controller is placed between the QPU and the quantum network since in this scenario, a quantum processing layer will be needed.}
    \label{fig:network_control}
\end{figure}

\subsection{Control System Architectures for Distributed Quantum Computing}

In this section, we discuss two possible network architectures for distributed quantum computing control systems. The major difference between the two systems is centralization of the control. In the first system, we consider a distributed architecture with a centralized control. In the second, the control is split such that each QPU in the system has its own control. In this section, we describe these two systems in depth. In later sections, we go into detail regarding the communication requirements needed to run the systems and potential protocols to achieve them.

\subsubsection{Centralised-Controlled Distributed Quantum Systems}

The first distributed quantum computing scenario we consider is depicted in Figure \ref{fig:single_control}. This scenario is one where there is a single controller and the quantum hardware behaves only according to the instructions that are fed from this controller. The QPU systems are connected to the controller via classical network and further they are connected to each other both classically and quantumly -- so that they can generate entanglement amongst themselves. The main idea here is that the CPUs in the network have a static IP and can be accessed by the centralized control. The finer synchronization between the QPU nodes is delegated to the CPU controlling the FPGA layer of the QPU from the centralized control ahead of execution time. The CPUs control the FPGAs and the FPGAs communicate over fixed low-latency links. This latency can be accounted for for control instruction scheduling. At a small and medium size, this network scheme will be best suited, but when many nodes are added to the network, a system with a distributed control is better suited, which we discuss in the next subsection.

\subsubsection{Decentralised-Controlled Distributed Quantum Systems}

With a decentralised control system, the nodes in the network are no longer in a \enquote{master-slave} relationship because the hardware is no longer controlled by a single entity. Resources in this setting need to be requested from various parties and there is no guarantee that the requested resource will be available at the time of request. Access to the controllers is hidden behind a firewall and their IP, MAC, and inner network configuration is potentially not exposed. We assume that the QPUs are offered by various vendors that have agreed to offer a base set of services: They provide access to quantum hardware for a maximum amount of time per instruction set execution, they offer classical communication input and output to a pre-specified IP address where the communication stream is established prior to execution, and they allow for remote entanglement to be established between quantum devices on specified quantum hardware. In this case, the control information between QPUs is needed and we will need a low-latency protocol that works in the network layer so that the control messages can be routed.

\subsection{Distributed Quantum Algorithm Scheduling}

In order for networked quantum hardware to execute instructions in a synchronous fashion, a method of dictating to the devices when the instructions should be executed is needed. In this section, we propose a temporal operation schedule, that is, a schedule of the operations with precise timestamps for execution. These schedules can then be sent to each QPU in the network with a time to begin execution. Because quantum gates generally have an upper bound for how long they take to execute, we can use this information when generating the schedule. Here, we assume that all gates have a known execution time as well and that latency times for classical communication and entanglement generation are known. We formalize the problem as follows:

\begin{problem}[Distributed Quantum Algorithm Scheduling] Given a distributed circuit as a series of gate layers, where each layer contains a collection of gates to be applied on the qubits in the system, and the gate times (i.e. the amount of time it takes to perform the gate) of each gate for each QPU in the system, produce a temporal gate execution schedule such that the following constraints are obeyed:
\begin{enumerate}
    \item Sending and receiving classical communication or entanglement between two parties occurs at the same time for the sending and the receiving parties.
    \item One qubit operation occurs per time instance per qubit for the duration of the gate time (i.e. no overlapping gates).
    \item At the start of a controlled operation, both qubits need to be available to perform the control gate (i.e. one qubit cannot have a gate operation ongoing).
\end{enumerate}
At the start of the problem, it is assumed that all nodes in the distributed system have synchronized clocks. We assume routes for any multi-hop communication or entanglement generation is already established and is already calculated into the communication time bounds. Moreover, it is assumed that swap gates are not considered in the scheduling and are assumed included in the worst case gate times provided.  
\end{problem}

Comparing and creating a hybrid temporal planning approach with constraint programming for quantum circuit scheduling has been investigated in Ref. \cite{booth2018comparing} for the max-cut problem. The difference here is the level of compilation is deeper as they include swap operations since they limit to nearest-neighbour interactions between the qubits. A temporal planning and constraint programming approach is therefore sensible. Here, we do not enforce nearest-neighbour interaction, and assume this process is included in the worse case timings for two qubit gates for swapping qubits to their nearest neighbour if needed and remapping the index of the qubit so it does not have to be swapped back. We assume at the end of each layer of gates, each qubit will be free to be operated on and no swapping is needed and therefore do not use any constraint programming. 

The output of the schedule for each QPU will have the form of Table \ref{tab:cir_qpu_sche}. Table \ref{tab:cir_sche} is an intermediate schedule which is used before splitting the schedules for each QPU. For a list of all possible commands and their descriptions, see Appendix A. We approach this problem as follows. We start with high-level instructions which are entanglement generation, single qubit gates, classical communication, and control gates. We generate an instruction list using this gate set. We then take the high-level circuits and break them down into finer control instructions. Once the full schedule is created, we can split the instructions such that the instruction schedule is for a single QPU. The instruction sets can then be sent to their respective QPUs and the algorithm can start. In order to ensure gates are performed in the correct order, we layer the circuits as done in the previous section and schedule the circuits layer by layer, iteratively constructing a full schedule. In complete form, we propose Algorithm \ref{algo:gate_ex_sche}.

\begin{table}[ht]
    \centering
    \begin{tabular}{|l|c|c|}
        \hline
        \bf{Command} & \bf{QPUs} &\bf{Time}\\
         \hline
         $CONTROL[G, \text{qID}, \text{qID}]$ &  QPU1 & $T_1$ \\
         \hline
         $GEN\_ENT
         [\text{qID, qID}]$ & QPU1, QPU2& $T_2$ \\
         \hline
         $CLASSICAL[\text{cID}]$ & QPU3, QPU2  & $T_3$ \\
         \hline
         \multicolumn{1}{|c|}{$\vdots$} & \multicolumn{1}{c}{$\vdots$} & \multicolumn{1}{|c|}{$\vdots$} \\
         \hline
         $SINGLE[G, \text{qID}]$ & QPU5 & $T_n$ \\
         \hline
    \end{tabular}
    \caption{$T_i$ is the time to execute the particular gate. The first step of Algorithm \ref{algo:gate_ex_sche} generates a table of gates with QPU information before filtering the gates for each QPU for execution.}
    \label{tab:cir_sche}
\end{table}

\begin{table}[ht]
    \centering
    \begin{tabular}{|l|c|c|}
        \hline
        \bf{Command} & \bf{Time}\\
         \hline
         $CONTROL[G, \text{qID}, \text{qID}]$ & $T_1$ \\
         \hline
         $SEND\_ENT[QPU, \text{qID}]$ & $T_2$ \\
         \hline
         $REC\_CLA[QPU, \text{cID}]$ & $T_3$ \\
         \hline
         \multicolumn{1}{|c|}{$\vdots$} & \multicolumn{1}{c|}{$\vdots$} \\
         \hline
         $SINGLE[G, \text{qID}]$ & $T_m$ \\
         \hline
    \end{tabular}
    \caption{$T_i$ is the time to execute the particular gate. The output of Algorithm \ref{algo:gate_ex_sche} will generate a collection of schedules in this form for each QPU.}
    \label{tab:cir_qpu_sche}
\end{table}

\begin{algo}[ht]
\small
\caption{Distributed Scheduler}
\textbf{Input:} 
\begin{compactitem}
    \item $QPUs$ the collection of QPUs in the system
    \item $C = \{l_1,..., l_n\}$ the circuit to schedule as a series of layers where each $l_i = \{g_{1},..., g_m\}$.
    \item $gateTime$ a mapping of gate names to time the gate takes to execute for each QPU.
\end{compactitem}
\textbf{Output:} A schedule of gate operations for each QPU to run.
\begin{algorithmic}[1]
    \State $layerEndtime \gets 0$
    \State $gateSchedule \gets [\hspace{1mm}]$
    \For{$l_i \in C$}\Comment{Make a first pass schedule based on the layers of the circuit}
        \For {$g_j \in l_i$}
            \State $gateSchedule.\textbf{add}((g_j, \textbf{QPUs}(g_j), layerEndtime))$
        \EndFor
        \State $layerEndtime\gets \max_j (gateTime(g_j)) + layerEndtime$
    \EndFor
    \State $QPUSchedules \gets \{\}$
    \For {$QPU \in QPUs$} \Comment{Split the schedules so each QPU has its own}
        \State $QPUSchedule \gets [\hspace{1mm}]$
        \For {$step \in gateSchedule$}
            \If {$QPU \in \textbf{QPUs}(step) \wedge  |\textbf{QPUs}(step)| = 1$}
                \State $QPUSchedule.\textbf{add}(step)$
            \ElsIf{$\textbf{gate}(step) = GEN\_ENT$}
                \If {$QPU = \textbf{QPUs}(step)[0]$} \Comment{Sender QPU}
                    \State $QPUSchedule.\textbf{add}(SEND\_ENT[\textbf{QPUs}(step)[1], \text{qID}], \textbf{time}(step))$
                \Else \Comment{Receiver QPU}
                \State $QPUSchedule.\textbf{add}(REC\_ENT[\textbf{QPUs}(step)[0], \text{qID}], \textbf{time}(step))$
                \EndIf
            \Else \Comment{The other non-local gate is classical transmission}
                \If {$QPU = \textbf{QPUs}(step)[0]$} \Comment{Sender QPU}
                \State $QPUSchedule.\textbf{add}(SEND\_CLA[\textbf{QPUs}(step)[1], \text{cID}], \textbf{time}(step))$
                \Else \Comment{Receiver QPU}
                \State $QPUSchedule.\textbf{add}(REC\_CLA[\textbf{QPUs}(step)[0], \text{cID}], \textbf{time}(step))$
                \EndIf
            \EndIf
        \EndFor
        \State $QPUSchedules[QPU]\gets QPUSchedule$
    \EndFor
    \State \textbf{return} $QPUSchedules$
\end{algorithmic}
\label{algo:gate_ex_sche}
\end{algo}

\subsection{Protocols}\label{sec:protocols}

In order to run a distributed quantum algorithm with a distribute quantum computer using the architectures proposed in the previous section, certain communication requirements are needed to ensure execution is possible. Protocols for controlling networked systems exist in practice in the centralized and decentralized case, and we explore some examples of them in this section. 

The first requirement considered is the classical communication between the controllers and the QPUs. Here what is needed is a method for sending  the computation instructions to the QPUs which can be done at slower latency, as well as a method for sending low-latency control bits between the QPUs. We explore methods for achieving this in the two cases. Clock synchronization is a commonly used scheme in distributed control systems. We consider an example of architectures using clock synchronization on a large scale.  Lastly in the multi-vendor case, we discuss certification steps needed to ensure multiple vendors are able to execute distributed quantum algorithms in a cooperative way. 

Selecting the specific hardware that can execute these protocols is left to future work as tasks such as entanglement generation is still it a primitive state and may not exist to the extent we need for years to come. Also as qubit technologies improve, the need for as-low-as-possible latency could be loosened, and other protocols could be used in replacement. Here we explore examples that could potentially achieve what is needed to perform distributed quantum computing.

\subsubsection{Classical Communication}

In order to run distributed quantum algorithms, there are specific non-local tasks that need to be carried out by the distributed system such as receiving control commands, sending measurement results to the controller, and sending qubit measurements between the QPUs at low latency to perform non-local control gates. In this section, we explore communication protocols which can be used by the control system to accomplish running distributed algorithms. Here we explore some examples of existing protocols that exist at an industry level. 

For the centralized control case, we neglect routing of information and assume each node is connected both classically and quantumly to another. As discussed,  We propose that there is a classical network connecting the QPUs to a centralized controller forming a \enquote{master-slave} relationship with the additional network of dedicated connections between the QPUs for the sole purpose of low-latency communication. This communication does not go through the CPU of the QPU, but directly between the FPGAs to perform the non-local gates.

When using a centralized control system, to perform slower communications between the QPUs and the controller one could consider two options. The first is to simply connect the CPU portions of he network to the controller using a local area network, and communicate using TCP/IP to from the controller to the QPUs. In this case one would need to closely monitor that communication traffic does not overwhelm the network. Another approach that has this feature built in is to use a protocol often used in industrial control systems. The Modbus communication protocol \cite{swales1999open}. Modbus is an open protocol used in a centralized controller master-slave setting as is this centralized controller setting. It is a messaging structure that allows for heterogeneous devices to communicate with a centralized controller and to receive control messages from the controller. The Modbus protocol can be used over a local network using TCP/IP making it easier to install into existing commonly used Ethernet networks. With Modbus, the controller can send the control instructions to the CPU portion of each QPU which can be sent to the FPGA to perform the portion of the quantum algorithm. With Modbus, the controller can also receive the qubit measurement results from the QPUs once the algorithm is complete. 

For low latency communication of short messages ($<$ 1 byte) between FPGAs there are existing methods that can be used to communicate at the ultra-low latency range ($<$ 1 ms). In the high performance computing domain, FPGA networks for ultra-low latency, high bandwidth communication are realized. Here we need to consider that the FPGAs may be meters apart. Connecting the FPGAs with, for example, 10 Gigabit Ethernet for sending short messages and using custom communication protocol and small form-factor pluggable (i.e. SFP+) transceivers, latency of 300 nanoseconds is possible for each link \cite{whichnas_2020}. With fibre, the latency can be even further reduced.

Another approach that can be integrated again comes from the industrial control domain. Industrial control, especially in the power sector faces issues where some devices need constant monitoring and reacting to the changes needs to be done at very fast speeds. A method used is the Mirrored Bits \cite{behrendt2001implementing} protocol. Mirrored Bits is a communication protocol for ultra-low latency communication adding additional a latency of approximately 200 $\mu$s for message processing in addition to the latency from transmitting signals over the communication link. Mirrored Bits could be used in this setting to transmit qubit measurement data. The devices that perform the Mirrored Bits protocol which are manufactured by Schweitzer Engineering Laboratories are programmable and can trigger different routines on the FPGA depending on the input bit. These devices are commonly used to frequently monitor sensor data to trigger emergency shut offs as fast as possible, for example. 

In the decentralized case, a dedicated wide area network could be established between the vendors such that a link-layer (of the OSI model) protocol is used for the classical control information between the QPUs. Low latency communication can be achieved using the link-layer protocol called the Generic Object Oriented Substation Event (GOOSE). In particular, IEC 61850 is a GOOSE Ethernet protocol meeting time sensitive communications and high-speed performance requirements of automation applications. At the link-layer over an Ethernet network experimental results show GOOSE can be used to transmit in the 0.5 ms scale \cite{youssef2020data}. When routing is involved, naturally, the latency will grow. If TCP/IP protocols are used over the Internet are considered, then it is unlikely one could create any latency guarantees. If instead there are dedicated wide area networks with routing, one could consider the network-layer version of GOOSE called Routable GOOSE \cite{apostolov2017r}. In \cite{youssef2020data}, R-GOOSE is analysed over a wide area network using a particular data distribution service and was shown to transmit with average latency of around 8 ms. 

Overall, there can be much to learn from looking into the power automation industry, as many low-latency and fast reaction systems have been developed which have carryover into distributed quantum computing. These protocols have been tested for robustness and security and could potentially fit well for doing distributed quantum computing over a wide area network. Moreover, networking FPGAs 

\subsubsection{Clock Synchronization}

For centralized control, the assumptions of clock synchronization and full connectivity at a small scale are not overly restrictive. High precision clock synchronization can be achieved even in very large configurations (i.e. that comprise a large number of devices) using methods such as in the White Rabbit Project \cite{serrano2013white}. White Rabbit is used at CERN to synchronize over 1000 nodes with sub-nanosecond accuracy. This is achieved using Ethernet with lengths of up to 10 km, with experiments demonstrating an average of 160 ps skew between similar clocks -- regarding clock environmental variables such as temperature -- after several hours \cite{wlostowski2011precise}. This protocol can be integrated in the centralized controller case so that all of the hardware used has synchronized clocks. Once the number of nodes in the network becomes large, routing and efficient network topologies becomes critical. 

In the decentralized case, the controllers  will need to perform a coarse-grain time synchronization via classical network synchronization protocols and a fine-grain synchronization, a precise notation of time can be shared. GPSs directly connected to FPGAs can be a solution where shielding does not stop the incoming signals. This process is common in distributed physical experiments such as in the Super-Kamiokande Detector \cite{fukuda2003super} and the CERN-OPERA experiment \cite{contaldi2011opera}. Each node will need to implement extra steps to guarantee that the timing information is constantly accurate.

\subsubsection{Entanglement Generation}

To perform the non-local control gates needed in the distributed circuits, the ability to share high quality entanglement between quantum processors is critical. Entangle generation has been achieved in various qubit technologies such as in optical photons, NV-centres, and super conducting qubits \cite{sanders2012review}, but entanglement generation in quantum networks is an ongoing research topic. We consider deterministic entanglement generation schemes such that there is a guaranteed entangled pair available shared between the quantum processors when it is needed. Experimental results demonstrating deterministic delivery of entanglement using NV-centers in diamond as qubits have been shown in  \cite{humphreys2018deterministic}, generating heralded entanglement at a rate of 39 hertz, three orders of magnitude better than previous known results and guaranteeing an entangled pair every 100 milliseconds with fidelity greater than 0.5 without pre- or post-selection. Methods for improving the results further are also proposed. This gives evidence that using entanglement to perform distributed quantum computing can become more feasible using various qubit technologies. As technology regarding entanglement generation and qubit stability improves, the deterministic entanglement generation rate can be improved. 

\begin{protocol}[ht]
\caption{Entanglement Validation}
\textbf{Vendor 1}
\begin{algorithmic}[1]
     \State Generate $N$ entangled pairs with Vendor 2
     \State Measure all of the owned halves
     \State Send $t < N$ randomly selected measurements to Vendor 2 without stating which qubits were measured
     \State Receive $t$ bits from Vendor 2
     \State If $t$ bits not received, abort
     \State Send positions of measurements to Vendor 2
     \State Receive positions of measurements from Vendor 2 and compare measurements
     \State Send acknowledgement if comparison passes, else send negative acknowledgement
\end{algorithmic}
\textbf{Vendor 2}
\begin{algorithmic}[1]
    \State Generate $N$ entangled pairs with Vendor 1
    \State Measure all of the owned halves
    \State Send $t < N$ randomly selected measurements to Vendor 1 without stating which qubits were measured
    \State Receive $t$ bits from Vendor 1
    \State If $t$ bits not received, abort
    \State Send positions of measurements to Vendor 1
    \State Receive positions of measurements from Vendor 1 and compare measurements
    \State Send acknowledgement if comparison passes, else send negative acknowledgement
\end{algorithmic}
\label{proto:ent_validation}
\end{protocol}

\subsubsection{Customer-Vendor Certification}

In the case of a decentralized controller, additional protocols are required to ensure that the all parties are able to execute the distributed quantum algorithms and an execution schedule can be made such that non-local operations are performed synchronously. In this setting, the user has no control over the quantum hardware and therefore a protocol for ensuring the user's instructions can be executed is needed. We write in Protocol \ref{proto:cert} a protocol for creating contracts between vendors and the user to ensure the desired instructions are carried out as specified. 

\begin{protocol}[ht]
\caption{Contract Creation Protocol}
\textbf{User}
\begin{algorithmic}[1]
  \State Assume it is known how many qubits exist on each available QPU for each QPU provider
  \State Generate non-local circuits Section \ref{sec:avqe}
  \State Request gate and classical communication latency times of gates from each QPU provider
  \State Generate a gate execution schedule using Algorithm \ref{algo:gate_ex_sche}
  \State Send executions schedules for the respective QPUs to the respective vendor along with current system time
  \State Await confirmation messages from all vendors
  \State If any vendor responds negatively, broadcast abort
  \State Gather all latest start times and broadcast start time as the minimum over all latest start times
  \State Await acknowledgements from all vendors, broadcast abort if any do not arrive, otherwise broadcast start signal
 \end{algorithmic}
\textbf{Vendor}
\begin{algorithmic}[1]
  \State Await request from user for gate times and respond accordingly
  \State Await gate execution schedule
  \State Validate that instructions can execute within allotted time frame for the user, respond to user if negative 
  \State If there are instructions with classical communication to an IP address, perform a handshake with other IP, respond to user if negative
    \State If there are instructions with entanglement, perform entanglement validation procedure in Protocol \ref{proto:ent_validation}, respond to user if negative 
    \State With other IPs, perform clock synchronization, respond to user if negative
    \State When all checks pass, respond positively to user with latest possible start time of execution adjusted for user system time difference
    \State Await start time confirmation, send acknowledgement to user, and await for final acknowledgement from user
 \end{algorithmic}
 \label{proto:cert}
\end{protocol}

\subsection{Deltaflow as a Networked Control System}

To orchestrate a centralized control system, we propose a scheme that uses Deltaflow.OS as the control. Deltaflow is based on the dataflow programming paradigm described in Section \ref{sec:dataflow}. The Deltaflow language is a Python based language that allows users to specify graphs and edges representing the dataflow between them. It is a hosted domain-specific language: the nodes are filled with code corresponding to the hardware that is represented. A Deltaflow OS is a tool for running Deltaflow programs on a specific piece of hardware. When given a Deltaflow program it performs compilation steps to transform it into chunks that run on the native hardware, runs those chunks, and exists in parallel to provide system services, networking and communication abstractions, and time sharing facilities. A Deltaflow OS provides the same functionality as an OS and compiler combination like Linux+GCC \cite{riverlane_2020}.

\subsubsection{Centralized Control}

\begin{figure}[ht]
    \centering
    \begin{tikzpicture}[scale=.9, every node/.style={transform shape}, 
    cpu/.style={scale=0.8, minimum size=0.75cm, draw=black, thick},
    node/.style={rounded corners, scale=0.8, minimum size=0.75cm, draw=black, thick},
  qpu/.pic = {
      \draw[line width=.4mm] (.2, 0.85) rectangle (4.3, 3.85);
      \foreach \x in {1, 2.25, 3.5} \foreach \y in {0.85, 1.70, 2.55}
      \draw[fill] (\x, 0.65 + \y) circle (2mm);
  }]
    \node (user) at (0, 1) { \Strichmaxerl[4] };
    \node[cpu] (user_cpu) at (0, 0) {User CPU};
    
    \node[node] (con_1) at (0, -1.25) {Controller};
    
    \node[node] (q_1) at (-3.5, -4.75) {Quantum Gates};
    \node[node] (c_1) at (-3.5, -2.5) {Classical Comm.};
    \node[node] (t_1) at (0, -3.625) {Time Ref.};
    \draw[scale=0.5] (-9.25, -15) pic {qpu};
    
    \node[node] (q_2) at (3.5, -4.75) {Quantum Gates};
    \node[node] (c_2) at (3.5, -2.5) {Classical Comm.};
    \draw[scale=0.5] (4.75, -15) pic {qpu};
    \draw[snake=coil, segment aspect=0, segment length=4pt] (-2.49, -6.25) -- (2.49, -6.25);
    \draw[<->] (user_cpu.south) -- (con_1.north);
    
    \draw[->] (t_1.north west) -- (c_1.south east);
    \draw[->] (t_1.north east) -- (c_2.south west);
    \draw[->] (t_1.south west) -- (q_1.north east);
    \draw[->] (t_1.south east) -- (q_2.north west);
    
    \draw[<->] (con_1.south west) -- (c_1.north east);
    \draw[<->] (con_1.south east) -- (c_2.north west);
    \draw[<->] ([xshift=1mm,yshift=-0.5mm]con_1.south west) -- ([xshift=-1mm,yshift=1mm]q_1.north east);
    \draw[<->] ([xshift=-1mm,yshift=-0.5mm]con_1.south east) -- ([xshift=1mm,yshift=1mm]q_2.north west);
    
    \draw[<->] (c_1.east) -- (c_2.west);
    \draw[<->] (c_1.south) -- (q_1.north);
    \draw[<->] (c_2.south) -- (q_2.north);
    \draw[<->] (q_1.south) -- (-3.5, -5.55);
    \draw[<->] (q_2.south) -- (3.5, -5.55);

    \draw[line width=.1mm, densely dashed, color=green!80!black] (c_1.north west)++(-0.2, 1.55) rectangle ([xshift=2mm,yshift=-2.5mm]q_2.south east);  
    \draw[line width=.1mm, densely dashed, color=red] (c_1.north west)++(-0.3, 0.3) rectangle ([xshift=3mm,yshift=-2.2cm]q_1.south east);  
    \draw[line width=.1mm, densely dashed, color=red] (c_2.north east)++(0.3, 0.3) rectangle ([xshift=-3mm,yshift=-2.2cm]q_2.south west);

    \node[scale=0.8] at (-3.85, -.9) {DeltaGraph};
    \node[scale=0.8] at (-4.5, -7.5) {QPU 1};
    \node[scale=0.8] at (4.5, -7.5) {QPU 2};
        
    \end{tikzpicture}
    \caption{A DeltaGraph laid out on top of distributed quantum hardware. The DeltaGraph controls the distributed computations performed by QPUs 1 and 2. The user enters the parameters to the centralized controller which initializes execution.}
    \label{fig:deltaflow_distributed_central}
\end{figure}
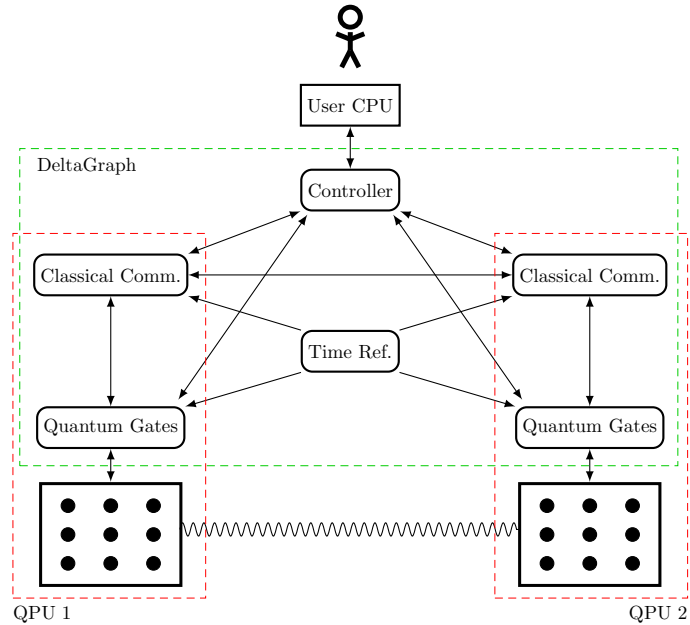

As discussed, Deltaflow uses nodes in a graph to control the flow of information within a hardware network. The nodes contain the logic of what instructions to perform when an input message is received. In this section, we describe the highest level nodes that are used and their instruction logic to conduct a distributed $\alpha$-VQE when there is a single centralized control. Within the high-level nodes can be more nodes but these nodes are hardware specific, and we leave this open for future work. In Figure \ref{fig:deltaflow_distributed_central}, we depict a Deltaflow graph laid on top of the hardware. There are four unique nodes in the DeltaGraph: Controller, Classical Communication, Quantum Gates, and Time Reference. We describe the controller responsibilities and the controlled node tasks below.

\textbf{Controller Node}:
The controller node is the main interface between the user and the distributed quantum hardware. This interaction closely resembles that of a distributed operating system as discussed in Section \ref{sec:dist_os}. The controller receives the algorithm parameters from the user, namely, the user sends the Ansatz preparation circuit designed for a single QPU and a Hamiltonian. Once received, the controller node handles the following:
\begin{enumerate}
    \item Takes user specified Ansatz preparation circuit and decomposed Hamiltonian, and confirms execution parameters
    \item When confirmed, the Hamiltonian is distributed across the QPUs and a gate execution schedule is made
    \item The schedule is split according to the locality of the execution, that is, the entire instruction set is not sent to each node, but rather just the parts that are executed by that particular node.
    \item Once all instructions are sent, the controller listens for incoming messages from any node in the DeltaGraph signaling an error. If such an error is received, then the execution process is aborted and the user is informed.
    \item Once scheduled run time has elapsed, the controller listens for measurement results from the QGNs and sends them to the user. 
\end{enumerate}

\textbf{Classical Communication Node (CCN)}: The CCN handles listening and sending classical data, that is information encoded into bits, to other QPUs in the network and collecting and forwarding classical control information from and to the quantum gates node (QGN) to perform any non-local control operations. The CCN receives the precise times for when to listen and when to transmit in the compiled instruction set from the controller. When time, the CCN communicates with the QGN to collect measurement results of the entangled qubits needed to perform the non-local gates and transmits the information to the paired QPU. The paired QPU will have its own CCN which will be listening for classical input and will know what to do with the information based on the predetermined instructions. Each CCN communicates with the controller to receive instructions at the beginning of execution. CCNs can also inform the controller of any failures in communication so that the controller can abort the run-time process.

\textbf{Quantum Gates Node (QGN)}:
The QGN controls the quantum hardware and controls the gate operations performed on and measurements of the qubits. It communicates to the CCN and the controller. The communication link with the CCN is used for sending the measurement results of the entangled qubits in the cat-entangler and cat-disentagler steps to the respectively paired QPU. With the controller, communication to receive instructions is needed as well as to transmit the qubit measurements at the end of the algorithm. Moreover, the QGN conducts entanglement half of an entanglement generation scheme, controlling the particular hardware as per the instructions and entanglement generation protocol selected.

\textbf{Time Reference Node (TRN)}: The TRN is a node used to maintain synchronous time within the DeltaGraph. The duty of the TNR is simply to periodically broadcast the system time so that all nodes in the network can update their own clocks to fix any clock drift that occurs during run time. Latencies from the TRN to the nodes can be accounted for, improving the overall resolution and precision of the timing reference.  

\subsubsection{Decentralized Control}

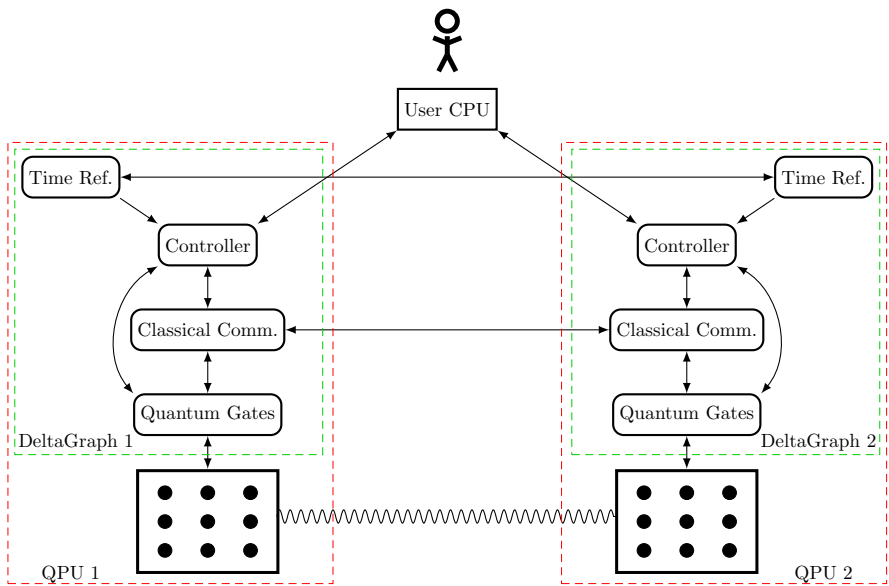
\begin{figure}[ht]
    \centering
    \begin{tikzpicture}[scale=0.9, every node/.style={transform shape}, 
    cpu/.style={scale=0.8, minimum size=0.75cm, draw=black, thick},
    node/.style={rounded corners, scale=0.8, minimum size=0.75cm, draw=black, thick},
  qpu/.pic = {
      \draw[line width=.4mm] (.2, 0.85) rectangle (4.3, 3.85);
      \foreach \x in {1, 2.25, 3.5} \foreach \y in {0.85, 1.70, 2.55}
      \draw[fill] (\x, 0.65 + \y) circle (2mm);
  }]
    \node (user) at (0, -0.25) { \Strichmaxerl[4] };
    \node[cpu] (user_cpu) at (0, -1.25) {User CPU};
    
    \node[node] (con_1) at (-3.5, -3.25) {Controller};
    \node[node] (q_1) at (-3.5, -5.75) {Quantum Gates};
    \node[node] (c_1) at (-3.5, -4.5) {Classical Comm.};
    \node[node] (t_1) at (-5.5, -2.25) {Time Ref.};
    \draw[scale=0.5] (-9.25, -17) pic {qpu};

    \node[node] (con_2) at (3.5, -3.25) {Controller};
    \node[node] (q_2) at (3.5, -5.75) {Quantum Gates};
    \node[node] (c_2) at (3.5, -4.5) {Classical Comm.};
    \node[node] (t_2) at (5.5, -2.25) {Time Ref.};
    \draw[scale=0.5] (4.75, -17) pic {qpu};
    
    \draw[->] (t_1.south east) -- (con_1.north west);
    \draw[->] (t_2.south west) -- (con_2.north east);
    
    \draw[<->] (user_cpu.south west) -- (con_1.north east);
    \draw[<->] (user_cpu.south east) -- (con_2.north west);
    
    \draw[<->] (c_1.east) -- (c_2.west);
    \draw[<->] (c_1.south) -- (q_1.north);
    \draw[<->] (c_2.south) -- (q_2.north);
    \draw[<->] (c_2.north) -- (con_2.south);
    \draw[<->] (c_1.north) -- (con_1.south);
    \draw[<->] (t_1.east) -- (t_2.west);
    \draw[<->] (q_1.south) -- (-3.5, -6.55);
    \draw[<->] (q_2.south) -- (3.5, -6.55);
    \draw (con_1.south west) edge[<->, out=210, in=130] (q_1.north west);
    \draw (con_2.south east) edge[<->, out=330, in=50] (q_2.north east);
    
    \draw[snake=coil, segment aspect=0, segment length=4pt] (-2.49, -7.25) -- (2.49, -7.25);
    
    \draw[line width=.1mm, densely dashed, color=green!80!black] (t_1.north west)++(-0.1, 0.1) rectangle ++(4.5, -4.5); 
    \draw[line width=.1mm, densely dashed, color=red] (t_1.north west)++(-0.2, 0.2) rectangle ++(4.75, -6.5); 
    
    \draw[line width=.1mm, densely dashed, color=green!80!black] (t_2.north east)++(0.1, 0.1) rectangle ++(-4.5, -4.5);   
    \draw[line width=.1mm, densely dashed, color=red] (t_2.north east)++(0.2, 0.2) rectangle ++(-4.75, -6.5); 
    
    \node[scale=0.8] at (-5.425, -6.15) {DeltaGraph 1};
    \node[scale=0.8] at (5.425, -6.15) {DeltaGraph 2};
    \node[scale=0.8] at (-5.5, -8.1) {QPU 1};
    \node[scale=0.8] at (5.5, -8.1) {QPU 2};
    
    \end{tikzpicture}
    \caption{A DeltaGraph for a decentralized control distributed quantum computer.}
    \label{fig:deltaflow_distributed_decentral}
\end{figure}

In the decentralized controller setting, the DeltaGraph is split between the QPUs in the network. Each QPU has their own DeltaGraph and the inter-communication is handled by the nodes in the DeltaGraph. In this case, there needs to be an agreed upon contract between each participating party which we have provided in Protocol \ref{proto:cert}. Here, for simplicity, we assume that there are no malicious parties and no eavesdroppers. Unlike the centralized case in the previous section, each remote QPU has its own controller and time reference node. The DeltaGraph in this case is depicted in Figure \ref{fig:deltaflow_distributed_decentral}. The main differences are that the independent time reference nodes allows different clock hardware for each vendor. Different mechanisms could be in place to adjust the time at each vendor which is controlled by the respective time reference node. Time reference nodes are connected such that at each vendor they can perform different pre-selected clock synchronization protocols. The controller nodes' main tasks are interfacing with the user and orchestrating the instructions provided by the user for the specific hardware of the vendor.  The controller for each vendor interfaces with the user similarly how it is described in the alternative method of deploying a distributed operating system in Section \ref{sec:dist_os}, where the user is aware of the network topology and hardware capabilities of each node. 

\textbf{Controller Node}:  
Each controller node has the responsibility of interfacing with the user and distributing the execution instructions to the other nodes. Much of the responsibilities of the controller in this case match that of the centralized controller case, but here the controller is only aware of the instructions that occur on the hardware in one QPU stack rather than the entire instruction set. In the centralized case, it was the duty of the controller to map the Ansatz circuit and schedule the Ans\"{a}zte to the distributed quantum hardware. In this case, that duty is moved to the user. The user queries QPU vendors, gathering the necessary hardware parameters in order to create a schedule that is executed on each vendor's hardware. The point of this is that the user will have more control over how many qubits they wish to run on each vendor's hardware. Variables such as cost of execution, time duration of execution, and vendor availability can be integrated into the user's decision when distributing their circuits across multiple vendors. The controller here can choose to reject or accept instruction sets depending on the vendor's hardware and availability. We summarize the responsibilities of the controller as follows: 
\begin{enumerate}
    \item Responds to user queries regarding hardware capabilities of the QPU as described in Protocol \ref{proto:cert}, orchestrating the handshake steps for classical communication and entanglement distribution.
    \item Once start time and instructions are gathered, the instructions are distributed to the respective nodes in the DeltaGraph.
    \item Time reference is sent to the nodes in the DeltaGraph via the controller in this case to simplify physical connections to any external clocks.
    \item Transmitting measurement results to back the user.
\end{enumerate}

\textbf{Classical Communications Node (CCN)}: At a high-level, the CCN performs much the same as in the centralized controller case. There may be added complexity in the low-level details depending on the selected protocols used for communication.

\textbf{Quantum Gates Node (QGN)}: The QGN performs the much of same tasks as in the centralized case. A difference to consider is that each vendor could run their own centralized control distributed quantum system. Here the quantum gate controller may need to be modified to perform much of the tasks of the centralized control as in the previous section.

\textbf{Time Reference Node (TRN)}: The TRN is distinct for each QPU, different than in the centralized case where there is only one TRN. Here the TRNs communicate with each other between DeltaGraphs in order to maintain clock synchronization in their respective graphs.

\section{Conclusions and Outlook}

In summary, we have explored how the generalized VQE algorithm $\alpha$-VQE can be distributed across arbitrary sized quantum computers connected with entanglement and classical communication networks. We proposed various approaches for splitting the Ansatz across the distributed system and provided algorithms for splitting the circuitry needed to split perform $\alpha$-QPE, the central algorithm around $\alpha$-VQE. We show in our analysis that with this approach, larger Ansatz stats used on distributed systems at the cost of run time. Next, we explore how one could network together a distributed quantum computer using two different architectures and we collect the necessary protocols needed to achieve this. We finish with a network control proposal using the Deltaflow.OS, the software based control system for controlling quantum systems at the various classical and quantum hardware levels.

What remains open is to implement this system at a small scale, using both simulation and physical systems. Initial effort has been made in this direction, but more work is needed to have a complete proof of concept. Already, an implementation using quantum network simulator QuNetSim \cite{diadamo2020qunetsim} and Deltaflow together has shown distributed quantum computing can be simulated. More, Deltaflow has been shown to work with distributed circuit boards which leads to the first steps of simulating distributed quantum algorithms. The methods for distributing the Ansatz states could consider more parameters for further optimization, such as the coherence times of the qubits and location of the control gates in the circuits. We aim to explore this in more detail in future work.

As discussed, distributed quantum computing is a promising path to developing large scale quantum computers. Much effort has gone into this in the classical computing domain, and the overlap between the fields is high. We can use this knowledge to design robust and secure distributed quantum computers, and as quantum technologies improve, this will surely become a reality.

\clearpage
\bibliographystyle{IEEEtran}

\begin{thebibliography}{10}
\providecommand{\url}[1]{#1}
\csname url@samestyle\endcsname
\providecommand{\newblock}{\relax}
\providecommand{\bibinfo}[2]{#2}
\providecommand{\BIBentrySTDinterwordspacing}{\spaceskip=0pt\relax}
\providecommand{\BIBentryALTinterwordstretchfactor}{4}
\providecommand{\BIBentryALTinterwordspacing}{\spaceskip=\fontdimen2\font plus
\BIBentryALTinterwordstretchfactor\fontdimen3\font minus
  \fontdimen4\font\relax}
\providecommand{\BIBforeignlanguage}[2]{{%
\expandafter\ifx\csname l@#1\endcsname\relax
\typeout{** WARNING: IEEEtran.bst: No hyphenation pattern has been}%
\typeout{** loaded for the language `#1'. Using the pattern for}%
\typeout{** the default language instead.}%
\else
\language=\csname l@#1\endcsname
\fi
#2}}
\providecommand{\BIBdecl}{\relax}
\BIBdecl

\bibitem{gambetta_2020}
\BIBentryALTinterwordspacing
J.~Gambetta, ``Ibm's roadmap for scaling quantum technology,'' Sep 2020.
  [Online]. Available:
  \url{https://www.ibm.com/blogs/research/2020/09/ibm-quantum-roadmap/}
\BIBentrySTDinterwordspacing

\bibitem{peruzzo2014variational}
A.~Peruzzo, J.~McClean, P.~Shadbolt, M.-H. Yung, X.-Q. Zhou, P.~J. Love,
  A.~Aspuru-Guzik, and J.~L. O’brien, ``A variational eigenvalue solver on a
  photonic quantum processor,'' \emph{Nature communications}, vol.~5, p. 4213,
  2014.

\bibitem{arute2020hartree}
F.~Arute, K.~Arya, R.~Babbush, D.~Bacon, J.~C. Bardin, R.~Barends, S.~Boixo,
  M.~Broughton, B.~B. Buckley, D.~A. Buell \emph{et~al.}, ``Hartree-fock on a
  superconducting qubit quantum computer,'' \emph{arXiv preprint
  arXiv:2004.04174}, 2020.

\bibitem{wang2019accelerated}
D.~Wang, O.~Higgott, and S.~Brierley, ``Accelerated variational quantum
  eigensolver,'' \emph{Physical review letters}, vol. 122, no.~14, p. 140504,
  2019.

\bibitem{yimsiriwattana2004generalized}
A.~Yimsiriwattana and S.~J. Lomonaco~Jr, ``Generalized ghz states and
  distributed quantum computing,'' \emph{arXiv preprint quant-ph/0402148},
  2004.

\bibitem{eisert2000optimal}
J.~Eisert, K.~Jacobs, P.~Papadopoulos, and M.~B. Plenio, ``Optimal local
  implementation of nonlocal quantum gates,'' \emph{Physical Review A},
  vol.~62, no.~5, p. 052317, 2000.

\bibitem{yimsiriwattana2004distributed}
A.~Yimsiriwattana and S.~J. Lomonaco~Jr, ``Distributed quantum computing: A
  distributed shor algorithm,'' in \emph{Quantum Information and Computation
  II}, vol. 5436.\hskip 1em plus 0.5em minus 0.4em\relax International Society
  for Optics and Photonics, 2004, pp. 360--372.

\bibitem{tannu2017taming}
S.~S. Tannu, Z.~A. Myers, P.~J. Nair, D.~M. Carmean, and M.~K. Qureshi,
  ``Taming the instruction bandwidth of quantum computers via hardware-managed
  error correction,'' in \emph{Proceedings of the 50th Annual IEEE/ACM
  International Symposium on Microarchitecture}, 2017, pp. 679--691.

\bibitem{fu2016heterogeneous}
X.~Fu, L.~Riesebos, L.~Lao, C.~G. Almudever, F.~Sebastiano, R.~Versluis,
  E.~Charbon, and K.~Bertels, ``A heterogeneous quantum computer
  architecture,'' in \emph{Proceedings of the ACM International Conference on
  Computing Frontiers}, 2016, pp. 323--330.

\bibitem{meter2006architecture}
R.~D.~V. Meter~III, ``Architecture of a quantum multicomputer optimized for
  shor's factoring algorithm,'' \emph{arXiv preprint quant-ph/0607065}, 2006.

\bibitem{cirac1999distributed}
J.~Cirac, A.~Ekert, S.~Huelga, and C.~Macchiavello, ``Distributed quantum
  computation over noisy channels,'' \emph{Physical Review A}, vol.~59, no.~6,
  p. 4249, 1999.

\bibitem{van2007communication}
R.~Van~Meter, K.~Nemoto, and W.~Munro, ``Communication links for distributed
  quantum computation,'' \emph{IEEE Transactions on Computers}, vol.~56,
  no.~12, pp. 1643--1653, 2007.

\bibitem{salamin2014schrodinger}
Y.~Salamin, ``Schr{\"o}dinger’s killer app: Race to build the world’s first
  quantum computer, by jonathan p. dowling: Scope: general interest. level:
  general readership,'' 2014.

\bibitem{hartnett2019new}
K.~Hartnett, ``A new law to describe quantum computing’s rise?'' 2019.

\bibitem{denchev2008distributed}
V.~S. Denchev and G.~Pandurangan, ``Distributed quantum computing: A new
  frontier in distributed systems or science fiction?'' \emph{ACM SIGACT News},
  vol.~39, no.~3, pp. 77--95, 2008.

\bibitem{meter2008arithmetic}
R.~V. Meter, W.~Munro, K.~Nemoto, and K.~M. Itoh, ``Arithmetic on a
  distributed-memory quantum multicomputer,'' \emph{ACM Journal on Emerging
  Technologies in Computing Systems (JETC)}, vol.~3, no.~4, pp. 1--23, 2008.

\bibitem{beals2013efficient}
R.~Beals, S.~Brierley, O.~Gray, A.~W. Harrow, S.~Kutin, N.~Linden, D.~Shepherd,
  and M.~Stather, ``Efficient distributed quantum computing,''
  \emph{Proceedings of the Royal Society A: Mathematical, Physical and
  Engineering Sciences}, vol. 469, no. 2153, p. 20120686, 2013.

\bibitem{cuomo2020towards}
D.~Cuomo, M.~Caleffi, and A.~S. Cacciapuoti, ``Towards a distributed quantum
  computing ecosystem,'' \emph{arXiv preprint arXiv:2002.11808}, 2020.

\bibitem{broadbent2009universal}
A.~Broadbent, J.~Fitzsimons, and E.~Kashefi, ``Universal blind quantum
  computation,'' in \emph{2009 50th Annual IEEE Symposium on Foundations of
  Computer Science}.\hskip 1em plus 0.5em minus 0.4em\relax IEEE, 2009, pp.
  517--526.

\bibitem{hornibrook2015cryogenic}
J.~Hornibrook, J.~Colless, I.~C. Lamb, S.~Pauka, H.~Lu, A.~Gossard, J.~Watson,
  G.~Gardner, S.~Fallahi, M.~Manfra \emph{et~al.}, ``Cryogenic control
  architecture for large-scale quantum computing,'' \emph{Physical Review
  Applied}, vol.~3, no.~2, p. 024010, 2015.

\bibitem{cruise2020practical}
J.~R. Cruise, N.~I. Gillespie, and B.~Reid, ``Practical quantum computing: The
  value of local computation,'' \emph{arXiv preprint arXiv:2009.08513}, 2020.

\bibitem{tanenbaum1995distributed}
A.~S. Tanenbaum, \emph{Distributed operating systems}.\hskip 1em plus 0.5em
  minus 0.4em\relax Pearson Education India, 1995.

\bibitem{tanenbaum2015modern}
A.~S. Tanenbaum and H.~Bos, \emph{Modern operating systems}, 2nd~ed.\hskip 1em
  plus 0.5em minus 0.4em\relax Pearson, 2001.

\bibitem{tanenbaum2007distributed}
A.~S. Tanenbaum and M.~Van~Steen, \emph{Distributed systems: principles and
  paradigms}, 2nd~ed.\hskip 1em plus 0.5em minus 0.4em\relax Prentice-Hall,
  2007.

\bibitem{johnston2004advances}
W.~M. Johnston, J.~P. Hanna, and R.~J. Millar, ``Advances in dataflow
  programming languages,'' \emph{ACM computing surveys (CSUR)}, vol.~36, no.~1,
  pp. 1--34, 2004.

\bibitem{peruzzo2013variational}
A.~Peruzzo \emph{et~al.}, ``A variational eigenvalue solver on a quantum
  processor. eprint,'' \emph{arXiv preprint arXiv:1304.3061}, 2013.

\bibitem{romero2018strategies}
J.~Romero, R.~Babbush, J.~R. McClean, C.~Hempel, P.~J. Love, and
  A.~Aspuru-Guzik, ``Strategies for quantum computing molecular energies using
  the unitary coupled cluster ansatz,'' \emph{Quantum Science and Technology},
  vol.~4, no.~1, p. 014008, 2018.

\bibitem{neumann2020imperfect}
N.~M. Neumann, R.~van Houte, and T.~Attema, ``Imperfect distributed quantum
  phase estimation,'' in \emph{International Conference on Computational
  Science}.\hskip 1em plus 0.5em minus 0.4em\relax Springer, 2020, pp.
  605--615.

\bibitem{kitaev1997quantum}
A.~Y. Kitaev, ``Quantum computations: algorithms and error correction,''
  \emph{Russian Mathematical Surveys}, vol.~52, no.~6, p. 1191, 1997.

\bibitem{saeedi2013linear}
M.~Saeedi and M.~Pedram, ``Linear-depth quantum circuits for n-qubit toffoli
  gates with no ancilla,'' \emph{Physical Review A}, vol.~87, no.~6, p. 062318,
  2013.

\bibitem{he2017decompositions}
Y.~He, M.-X. Luo, E.~Zhang, H.-K. Wang, and X.-F. Wang, ``Decompositions of
  n-qubit toffoli gates with linear circuit complexity,'' \emph{International
  Journal of Theoretical Physics}, vol.~56, no.~7, pp. 2350--2361, 2017.

\bibitem{maslov2016advantages}
D.~Maslov, ``Advantages of using relative-phase toffoli gates with an
  application to multiple control toffoli optimization,'' \emph{Physical Review
  A}, vol.~93, no.~2, p. 022311, 2016.

\bibitem{babbush2018low}
R.~Babbush, N.~Wiebe, J.~McClean, J.~McClain, H.~Neven, and G.~K.-L. Chan,
  ``Low-depth quantum simulation of materials,'' \emph{Physical Review X},
  vol.~8, no.~1, p. 011044, 2018.

\bibitem{michielsen2017benchmarking}
K.~Michielsen, M.~Nocon, D.~Willsch, F.~Jin, T.~Lippert, and H.~De~Raedt,
  ``Benchmarking gate-based quantum computers,'' \emph{Computer Physics
  Communications}, vol. 220, pp. 44--55, 2017.

\bibitem{seeley2012bravyi}
J.~T. Seeley, M.~J. Richard, and P.~J. Love, ``The bravyi-kitaev transformation
  for quantum computation of electronic structure,'' \emph{The Journal of
  chemical physics}, vol. 137, no.~22, p. 224109, 2012.

\bibitem{bergholm2018pennylane}
V.~Bergholm, J.~Izaac, M.~Schuld, C.~Gogolin, C.~Blank, K.~McKiernan, and
  N.~Killoran, ``Pennylane: Automatic differentiation of hybrid
  quantum-classical computations,'' \emph{arXiv preprint arXiv:1811.04968},
  2018.

\bibitem{hespanha2007survey}
J.~P. Hespanha, P.~Naghshtabrizi, and Y.~Xu, ``A survey of recent results in
  networked control systems,'' \emph{Proceedings of the IEEE}, vol.~95, no.~1,
  pp. 138--162, 2007.

\bibitem{yang2006networked}
T.~C. Yang, ``Networked control system: a brief survey,'' \emph{IEE
  Proceedings-Control Theory and Applications}, vol. 153, no.~4, pp. 403--412,
  2006.

\bibitem{mahmoud2019networked}
M.~S. Mahmoud and Y.~Xia, \emph{Networked control systems: cloud control and
  secure control}.\hskip 1em plus 0.5em minus 0.4em\relax
  Butterworth-Heinemann, 2019.

\bibitem{booth2018comparing}
K.~E. Booth, M.~Do, J.~C. Beck, E.~Rieffel, D.~Venturelli, and J.~Frank,
  ``Comparing and integrating constraint programming and temporal planning for
  quantum circuit compilation,'' \emph{arXiv preprint arXiv:1803.06775}, 2018.

\bibitem{swales1999open}
A.~Swales \emph{et~al.}, ``Open modbus/tcp specification,'' \emph{Schneider
  Electric}, vol.~29, 1999.

\bibitem{whichnas_2020}
\BIBentryALTinterwordspacing
WhichNAS, ``10gbase-t vs. sfp+ - which is the best 10g network solution for
  small businesses,'' Jun 2020. [Online]. Available:
  \url{https://whichnas.com/10gbase-t-vs-sfp-which-is-the-best-10g-network-solution-for-small-businesses/}
\BIBentrySTDinterwordspacing

\bibitem{behrendt2001implementing}
K.~Behrendt and K.~Fodero, ``Implementing mirrored bits technology over various
  communications media,'' \emph{SEL Application Guide}, vol.~12, 2001.

\bibitem{youssef2020data}
T.~A. Youssef, M.~M. Esfahani, and O.~Mohammed, ``Data-centric communication
  framework for multicast iec 61850 routable goose messages over the wan in
  modern power systems,'' \emph{Applied Sciences}, vol.~10, no.~3, p. 848,
  2020.

\bibitem{apostolov2017r}
A.~Apostolov, ``R-goose: what it is and its application in distribution
  automation,'' \emph{CIRED-Open Access Proceedings Journal}, vol. 2017, no.~1,
  pp. 1438--1441, 2017.

\bibitem{serrano2013white}
M.~Lipi{\'n}ski, T.~W{\l}ostowski, J.~Serrano, and P.~Alvarez, ``White rabbit:
  A ptp application for robust sub-nanosecond synchronization,'' in \emph{2011
  IEEE International Symposium on Precision Clock Synchronization for
  Measurement, Control and Communication}.\hskip 1em plus 0.5em minus
  0.4em\relax IEEE, 2011, pp. 25--30.

\bibitem{wlostowski2011precise}
T.~W{\l}ostowski, ``Precise time and frequency transfer in a white rabbit
  network,'' Ph.D. dissertation, Instytut Radioelektroniki, 2011.

\bibitem{fukuda2003super}
S.~Fukuda, Y.~Fukuda, T.~Hayakawa, E.~Ichihara, M.~Ishitsuka, Y.~Itow,
  T.~Kajita, J.~Kameda, K.~Kaneyuki, S.~Kasuga \emph{et~al.}, ``The
  super-kamiokande detector,'' \emph{Nuclear Instruments and Methods in Physics
  Research Section A: Accelerators, Spectrometers, Detectors and Associated
  Equipment}, vol. 501, no. 2-3, pp. 418--462, 2003.

\bibitem{contaldi2011opera}
C.~R. Contaldi, ``The opera neutrino velocity result and the synchronisation of
  clocks,'' \emph{arXiv preprint arXiv:1109.6160}, 2011.

\bibitem{sanders2012review}
B.~C. Sanders, ``Review of entangled coherent states,'' \emph{Journal of
  Physics A: Mathematical and Theoretical}, vol.~45, no.~24, p. 244002, 2012.

\bibitem{humphreys2018deterministic}
P.~C. Humphreys, N.~Kalb, J.~P. Morits, R.~N. Schouten, R.~F. Vermeulen, D.~J.
  Twitchen, M.~Markham, and R.~Hanson, ``Deterministic delivery of remote
  entanglement on a quantum network,'' \emph{Nature}, vol. 558, no. 7709, pp.
  268--273, 2018.

\bibitem{riverlane_2020}
\BIBentryALTinterwordspacing
Riverlane, ``Deltaflow®,'' Sep 2020. [Online]. Available:
  \url{https://www.riverlane.com/products/}
\BIBentrySTDinterwordspacing

\bibitem{diadamo2020qunetsim}
S.~DiAdamo, J.~N{\"o}tzel, B.~Zanger, and M.~M. Be{\c{s}}e, ``Qunetsim: A
  software framework for quantum networks,'' \emph{arXiv preprint
  arXiv:2003.06397v4}, 2020.

\end{thebibliography}


\appendix
\section{Additional Material}
\subsection{Algorithms}

\begin{algo}[H]
\small
\caption{Does Not Fit}
\textbf{Input:}
\begin{compactitem}
    \item $Q = [q_1, ..., q_n]$ the collection of QPUs in the distributed system, non-increasingly sorted by number of available qubits
    \item $A$ the size of the Ansatz
\end{compactitem}
\textbf{Output:} If the Ansatz can fit in the distributed QPU $Q$\\
\textbf{DoesNotFit}$(Q, A)$
\begin{algorithmic}
    \If{$Q$ is empty}
    \State \textbf{return} $true$
    \EndIf
    \For{$q_i \in Q$}
        \State $curAllocation \gets [0,..., 0]$ \Comment{Allocate $n$ element array of $0$s}
        \State $possibleQPUs \gets \{q_1,...,q_i\} \subseteq Q$
        \If {$i == 1$}
            \State $k \gets \textbf{QPUNumber}(possibleQPUs[1])$ \Comment{The QPU index}
            \State $curAllocation[k] \gets q_1 - 1$
        \Else
            \State State $k \gets \textbf{QPUNumber}(possibleQPUs[1])$ \Comment{The QPU index}
            \State $curAllocation[k] \gets q_1 - 3$
            \For{$q_j \in \{q_2, ..., q_i\}$}
                \State  $k \gets \textbf{QPUNumber}(possibleQPUs[j])$ \Comment{The QPU index}
                \State $curAllocation[k] \gets q_j - 2$
            \EndFor
        \EndIf
        \If{$\textbf{sum}(curAllocation) \geq A$}
            \State \textbf{return} $false$ \Comment{The Ansatz does fit}
        \EndIf
    \EndFor
     \State \textbf{return} $true$
\end{algorithmic}
\label{algo:does_not_fit}
\end{algo}

\subsection{Control System Commands}
\begin{table}[H]
    \centering
    \begin{tabularx}{\textwidth}{|s|L|}
        \hline
        \bf{Command}  &\bf{Description}\\
         \hline
         $TWO\_QUBIT[G, \text{qID}_1, \text{qID}_2]$ & Two qubit gate $G$ on qubits with memory IDs $\text{qID}_1, \text{qID}_2$, where $\text{qID}_1$ is the control if needed. $\text{qID}_1$ is potentially a classical register.  \\ 
         \hline
         $SINGLE[G, \text{qID}]$ & A single qubit gate $G$ applied to qubit with memory ID qID\\ 
         \hline
         $GEN\_ENT[\text{qID}_1, \text{qID}_2]$ & Generate entanglement and store it in the qIDs $\text{qID}_1, \text{qID}_2$\\ 
         \hline
         $REC\_ENT[QPU, \text{qID}]$ & Receive entanglement half and store it in at memory qID\\ 
         \hline
         $SEND\_ENT[QPU, \text{qID}]$ & Send entanglement half and keep other half in qubit memory qID \\ 
         \hline
         $CLASSICAL[\text{cID}]$ & Transmission of classical bit in classical memory with ID cID\\ 
         \hline
         $SEND\_CLA[QPU, \text{cID}]$ &  Send classical bit in classical memory with ID cID \\ 
         \hline
         $REC\_CLA[QPU, \text{cID}]$ &  Receive classical bit and store in classical memory with ID cID \\ 
         \hline
    \end{tabularx}
    \caption{Commands for QPU schedule}
    \label{tab:controls}
\end{table}

\end{document}